\documentclass{aa}
\usepackage{graphicx}
\usepackage{enumerate}
\usepackage{txfonts}
\usepackage{natbib}
\bibpunct{(}{)}{;}{a}{}{,}

\begin{document}

\title{The s-process in the Galactic halo:\\
the fifth signature of spinstars in the early Universe?}

\author {G. Cescutti\inst{1} \thanks {email to: cescutti@aip.de} \and
  C. Chiappini \inst{1} \and R. Hirschi \inst{2,3} \and G. Meynet
  \inst{4} \and U. Frischknecht \inst{2,5} }
\institute{Leibniz-Institut f\"ur Astrophysik Potsdam, An der
  Sternwarte 16, 14482, Potsdam, Germany 
\and Astrophysics Group, Lennard-Jones Laboratories, EPSAM, Keele University, ST5 5BG, Staffordshire, UK
\and Kavli Institute for the Physics and Mathematics of the Universe,
University of Tokyo, 5-1-5 Kashiwanoha, Kashiwa, 277-8583, Japan
\and Geneva Observatory, Geneva University, 1290 Sauverny, Switzerland
\and Department of Physics, University of Basel, Klingelbergstrasse 82, 4056 Basel, Switzerland}
\date{Received xxxx / Accepted xxxx}

\abstract {Very old halo stars have previously been found to show at least
  four different abundance anomalies, which models of fast-rotating
  massive stars (\emph{spinstars}) can successfully account for: rise
  of N/O and C/O, low $^{12}$C/$^{13}$C, and a primary-like evolution
  of Be and B.  Here we show the impact of these same stars in the
  enrichment of Sr and Ba in the early Universe.}  {We study whether
  the s-process production of fast-rotating massive stars can offer an
  explanation for the observed spread in [Sr/Ba] ratio in halo stars
  with metallicity [Fe/H]$< -$2.5.}  { By means of a chemical
  inhomogeneous model, we computed the enrichment of Sr and Ba by
  massive stars in the Galactic halo. Our model takes, for the first
  time, the contribution of spinstars into account.}  {The model
  (combining an r-process contribution with an s-process from
  fast-rotating massive stars) is able to reproduce the observed
  scatter in the [Sr/Ba] ratio at [Fe/H]$<-2.5$.  Toward higher
  metallicities, the stochasticity of the star formation fades away
  owing to the increasing number of exploding and enriching stars, and
  as a consequence the predicted scatter decreases.}  {Our scenario is
  again based on the existence of \emph{spinstars} in the early
  Universe.  Very old halo stars have previously been found to show at
  least four other abundance anomalies, which rotating models of
  massive stars can successfully account for. Our results provide a
  fifth independent signature of fast-rotating massive
  stars: an early enrichment of the Universe in s-process elements.}

\keywords{Galaxy: evolution -- Galaxy: halo -- 
stars: abundances -- stars: massive -- stars: rotation -- nuclear reactions, nucleosynthesis, abundances }

\titlerunning{The 5th signature of spinstars in the Galactic halo }

\authorrunning{Cescutti et al.}

\maketitle

\section{Introduction}

Soon after the Big Bang, the appearance of the first stellar
generations drastically changed the course of the history of the
Universe by enriching the primordial gas with elements heavier than
helium (referred to as metals) through both stellar winds and
supernova explosions. High-resolution hydrodynamical simulations of
the formation of the first stars suggest that these objects have formed
in dark matter mini-halos and  have played a key role in the
formation and properties of the first galaxies \citep[see][for a
review]{Karlsson11}.  One of the key questions that drive enormous
theoretical and observational efforts is whether the primordial
environment in which the first stars were born made them different
from present-day stars.  The answer to this question is crucial for
understanding the impact of the first stars, in terms both of metal
injection and energetic feedback, on the formation of the first
galaxies \citep{BrommY11}. Major endeavors, such as the James Webb
Space Telescope or the 40m European Extremely Large Telescope,
expected to detect the first galaxies, have been pushing the frontiers
of high-resolution hydro-dynamical simulations with the aim of predicting
their luminosities and colors \citep{JoggW11}.  These predictions,
however, strongly depend on the properties of the first stars,
including their mass spectrum and chemical composition.

To simulate the formation of the first stars is technically very
challenging, because one starts from large-scale cosmological
simulations ($> 10^{8}$pc) that are then zoomed into the highest
density peaks ($< 10^{-4}$pc), covering a dynamical range of at least
10$^{12}$. When the resolution of these simulations increases, the
conclusions in this (still emerging) field change drastically. The
best example is the predicted mass spectrum of the first stars: while
most of the theoretical simulations predicted that primordial stars
were predominantly massive \citep{Bromm99,Abel02,BrommL04}, current
results suggest that their mass is in the more normal range for
massive stars (between 10 and at most 50 M$_{\odot}$) and that even
low-mass stars could also have formed in the very early Universe
\citep{Stacy11,Stacy12,Clark11,Greif11,Greif12}.

These state-of-the art simulations have become sophisticated, and are
now able to resolve the accretion disk onto the protostellar
core. However, the effects of turbulence, magnetic fields and
  negative feedback by the forming protostar HII region are complex,
  leading to large model uncertainties. Moreover, an enormous
  resolution is needed to correctly follow the accretion
  process \citep[see][for a recent discussion]{Ferrara12}.  As a
consequence, the current mass spectrum of the first stars is still not
known. Very interestingly, the most recent simulations seem to
converge to one result: there is sufficient angular momentum in the
protocloud collapse to yield rapidly rotating stars near the break-up
speeds \citep[e.g.][]{Stacy11,Stacy12}. As we see below, the
completely independent approach of Galactic Archaeology, had already
pointed in the same direction.

It is now recognized that Galactic Archaeology offers a powerful
alternative way of probing the nature of the first stellar generations
\citep{Freeman,BrommY11,Tum10}, thus providing invaluable constraints
to the above-mentioned simulations. It is in the Milky Way and its
satellites that the oldest and most metal-poor stars in the Universe
are observable, because they were born at times or equivalent
redshifts that are still out of reach of even the deepest planned
extragalactic surveys.

Low-mass stars (masses below 80\% of a solar mass) have comparable
lifetimes to the age of the Universe. We, therefore, expect that in
their atmospheres, the elemental abundances of the gas at the time of
their birth are mostly preserved, offering a local benchmark to
cosmology. These stars thus keep memory of the unique nucleosynthesis
in the first stellar generations, thereby providing invaluable constraints on
the masses of the first stars, hence on their stellar yields. This is
the very basic input for modeling subsequent stellar generations,
leading to the abundance distribution prevailing at present time.

Reliable age determination, especially for the oldest stars, has been
one of the biggest challenges in astronomy, although a breakthrough is
expected to come from astroseismology \citep{Chiappini11as}.
In  some rare cases it was possible to estimate an age by measuring
  long-lived radioactive elements such as Th and U \citep{Cayrel01},
 albeit with still non-negligible uncertainties. In the
meantime, the strategy adopted so far has been to look for the most
metal-poor stars in the Milky Way and its satellites, both presumably
formed from a gas enriched only by a few first supernovae.

The hunt for the most metal-poor stars in the Universe is not an easy
task. This search is restricted to nearby (resolved) stellar
populations: the MW halo and its satellites. The chemically most
primitive stars are extremely rare. Despite their scarcity,
in the MW halo around 130 halo stars, with metallicities below [Fe/H]
= $-$3, are  currently known \citep[][for a recent review]{Frebel10}. Only two of
them have [Fe/H] $< -$5 \citep{Frebel05,Frebel08,Christlieb02}, while more
recently, \citet{Caffau11} has discovered a star with an [Fe/H] = $-$4.89
(so slightly above $-$5) but with the lowest global metallicity (Z$\le
7.4 \times 10^{-7})$ not enhanced in C and N differently from
the Frebel and the Christlieb stars.

Recently, another window into the very metal-poor universe has been
provided by the discovery of the so-called ultra-faint dwarf
spheroidal galaxies. These objects are dominated by dark matter and
have very low average metallicities (some show a mean [Fe/H] $\sim$
$-$2.6), thus offering an excellent environment in which to search for
more fossil records of the early enrichment.  The search for more of
these Rosetta stones in the history of the Universe constitutes the
goal of several planned and ongoing surveys, and for details the
interested reader is referred to SEGUE \citep{segue}, APOGEE
\citep{apogee}, LAMOST \citep{lamost}, SkyMapper \citep{skymapper},
4MOST \citep{4most}, and WEAVE \citep{weave}.

The new information on cosmic abundances provided by the still modest
samples of very metal-poor stars in the halo (still confined to mostly
nearby and brighter objects) and the MW satellites have already proven
to be a rich source of information on the nature of the first stellar
generations. In their pioneering work, \citet{Cayrel04} derived
abundances of several elements for a fairly large sample of very
metal-poor normal stars (not C-enhanced),
obtained with the ESO Large Program ``First Stars''. Their data of
unprecedented quality reveals a striking homogeneity in the chemical
properties of halo stars.

At first sight, this uniformity was interpreted as a challenge to the
view that the whole Galactic halo formed from the successive
swallowing of smaller stellar systems with independent evolutionary
histories, as predicted from the cold dark matter theory. The
situation turned out to be more complex, since the same stars were found
to show large scatter in r- and s-process elements \citep{Franc07} and
in the N/O ratios \citep{Spite05}. 

The first stars are believed to have been radically different from
present-day massive stars, because they were metal-free. The lack (or
only traces) of metals leads to faster surface rotation velocities, since
metal-poor stars are more compact than metal-rich ones. Stars formed
from a gas whose global metallicity is below $\sim$1/2000 that of the
Sun (i.e. Z = 10$^{-5}$) could attain rotational velocities of
500-800km/s \citep[depending on the stellar mass, hereafter called
\emph{spinstars} -- see][for a recent review]{MM12}.

Two important effects of interest for the present work  arise in
\emph{spinstars}: 
\begin{enumerate}[a)]
\item rotation triggers mixing processes inside the star, which leads to
producing large quantities of primary $^{14}$N, $^{13}$C and $^{22}$Ne
\citep[][and references therein]{Meynet08}. The production of primary
$^{22}$Ne has a strong impact on the s-process nucleosynthesis in
\emph{spinstars} compared to nonrotating stars, increasing the
s-process yields of heavy elements by orders of magnitude
\citep[e.g.][]{Pigna08,Chiappini11,Frisch12};
\item the strong mixing caused by rotation enriches the stellar surface,
thus increasing the opacity of the outer layer and producing
line-driven winds \citep[but see][]{Muy}.  
In addition, fast rotation can also lead to mechanical mass-loss. Both
mechanisms are already able to trigger non-negligible mass loss at
very low metallicities (a result not achieved by standard models
without rotation).
\end{enumerate}
 We have shown that a generation of \emph{spinstars} in the very early
Universe is currently the most promising 
  for the increasing of N/O ratio observed in very metal-poor normal stars
in the Galactic halo \citep{Chiappini06}. Indeed, asymptotic
  giant branch stars (AGBs), that produce $^{14}$N and $^{13}$C
  efficiently, have not had time to contribute to the chemical
  enrichment of the early Universe due to their long lifetimes;
another viable explanation not fully explored yet is the contribution of SAGB.
 Our models with \emph{spinstars} also naturally
account for the observed increase in the C/O ratio in halo stars
towards low metallicities \citep{Chiappini06,Fabbian}.
Moreover, if
\emph{spinstars} were common phenomena in the early Universe, the
metal-poor ISM of galaxies with a star formation history similar to
the one inferred for our Galactic halo, should have $^{12}$C/$^{13}$C
ratios between 30-300 below [Fe/H]=$-$3. Without fast rotators, the
predicted $^{12}$C/$^{13}$C ratios is 4,500 at [Fe/H]=$-$3.5
increasing to 31,000 at [Fe/H] = $-$5.0 \citep{Chiappini08}. Current
observations of this isotopic ratio in very metal-poor giant stars in
the Galactic halo \citep{Spite06} agree better with chemical
evolution models that include fast rotators. 
 Recently, the same results for C 
isotopes and for the ratio N/O have been obtained in \citet{Kobayashi11}
using another chemical model.

To test the last predictions, challenging measurements of the
$^{12}$C/$^{13}$C in more extremely metal-poor giants or turnoff stars
are required (only feasible with 30m-class telescopes). Furthermore,
we explain the apparently contradictory finding of a large scatter in
C, N and the almost complete lack of scatter in [$\alpha$/Fe] ratios
of the same stars as being the consequence of the combination of a
stochastic star formation rate in the early universe (hence,
incomplete sampling of the IMF and incomplete halo mixing), coupled
with a strong yield dependence on the stellar mass for these
particular elements \citep{Cesc10}: \emph{spinstars} provide a
nucleosynthetic site in which the yields of C, N strongly
depend on stellar mass.

In summary, we claim that the rise in N/O and C/O and the low
$^{12}$C/$^{13}$C in very metal-poor halo stars are three footprints
of the existence of \emph{spinstars} in the earliest phases of the
Universe. A fourth signature has recently been suggested by
\citet{Prantzos12} who argue that the observed primary-like evolution
of Be and B can be explained if galactic cosmic rays are accelerated
from the wind material of rotating massive stars, rich in CNO, hit by
the forward shock of the subsequent supernova explosions. These four
footprints have been found in the very metal-poor Universe ([Fe/H] $<
-$3).

In the present paper we suggest that two puzzling results involving
n-capture elements, such as Y, Sr, and Ba, might again point to the
existence of \emph{spinstars} in the early Universe, namely:

\begin{itemize}

\item The large scatter in the abundance ratios of s-process elements
  from the first and second peaks, e.g., [Sr/Ba] (see
  Section~\ref{sec:heavy}), which cannot be explained even by
  inhomogeneous models that otherwise match the lack of scatter in
  $\alpha$-elements, and the large scatter of the same elements with
  respect to iron \citep[e.g., Sr and Ba in][]{Cesc08};

\item The large enhancements of Ba, La, Y, and Sr found in stars of NGC
  6522, a bulge globular cluster shown to be the oldest 
one in the  Milky Way \citep{Barbuy09,Chiappini11}.

\end{itemize}

Indeed, \citet{Chiappini11} have shown that a) the scatter in the
[Sr/Ba] and [Y/Ba] for very metal-poor halo stars is similar to what is
found in the NGC 6522 stars, and b) that the enrichment of Sr, Y, Ba,
and La in five out of eight stars of NGC 6522 could be either the
s-process activation in early generations of \emph{spinstars} or the
s-process due to the contamination from a low-mass AGB stars in a
binary system, with initial metallicity coincident with the metal
content of the cluster. In the first case, the chemical
  enrichment is from stars polluting the primordial material before
  forming the cluster, whereas in the second the chemical
  enrichment would happen via AGB-mass transfer at whatever point in
  the history of NGC 6522, thus polluting the stars we are now
  observing.  However, the remaining three stars showing the highest
[Y/Ba] are not compatible with s-process nucleosynthesis in AGB stars
and can be readily explained from early enrichment from \emph{spinstars}.
It is now possible to investigate this hypothesis in a more
quantitative way by using inhomogenous chemical evolution models
computed with the most recent calculations carried out by our group
(Frischknecht et al. 2012, 
Frischknecht et al. in preparation). In the
present work, we focus on the observed scatter of [Sr/Fe],
[Ba/Fe], and [Sr/Ba] in very metal-poor halo stars. 
We argue that fast-rotating massive stars have also left a
footprint in the early enrichment of heavy elements, such as Sr and
Ba. In a forthcoming paper we will then specifically discuss models
for the Galactic bulge.

The paper is organized as follows. In Section 2 we describe the
current situation with respect to the predicted stellar yields of
heavy elements and justify the use of \emph{empirical yields} for the
r-process In Section 3 the observational data is described. Sections 4
and 5 describe the chemical model and the adopted stellar yields,
respectively. In Section 5 our results are presented, and in Section 6
we summarize our conclusions.

\section{The production of heavy elements: an open debate}
\label{sec:heavy}

Elements with Z $>$30 are labeled neutron capture elements: they are
mainly formed through neutron captures, and not through fusion,
because charged particle reactions on elements beyond iron (Z = 26)
are endothermic (omitting the p-process, responsible for less than 1\%
of the heavy elements) .  The neutron capture process is split in
rapid process (r-process) or slow process (s-process) depending on
whether the timescale for neutron capture $\tau_{n}$ is faster or
slower than radioactive beta decay $\tau_{\beta}$, according to the
initial definition by \citet{BBFH57}.

There are several sites of production for the s-process. Massive stars
produce the so-called weak s-process (60$<$A$<$90) and intermediate-
and low-mass stars produce the main s-process (up to lead) -- see
recent review by \citet{Kaeppeler}.  For the r-process, the final states
of massive stars, where extremely high neutron density are achievable,
seem to provide a promising site \citep[although still very uncertain,
see][]{Thielemann11}. A multiplicity of hypotheses does exist, such as
neutrino winds during SNII \citep{WWM94,TWJ94,Thom03,AJS07,Farouqi09},
neutron star mergers \citep{FRT99,Goriely11,Korobkin12}, O-Mg-Ne core
explosion \citep{WJM11}, asymmetric explosions \citep{Cam2003,AGT07},
quark novae \citep{JMO07}, and magnetorotationally driven supernovae
\citep{Winteler12}.

The general complexity of these environments, coupled with the
inaccessibility of the nuclear data of the isotopes involved, leads to
still uncertain predictions for the stellar yields.  Only recently have a
few SNII simulations started to successfully explode stellar
cores, and the results are still controversial.  This is the case for
the simulations of O-Mg-Ne core collapse supernovae (SN) by
\citet{KJH06} and the nucleosynthesis calculated on this basis by
  \citet{WJM11, WNJ09}.

  Even though the r-process is not clearly understood, most recent
  studies in this field point to massive stars (with the
    remarkable exception of the neutron star mergers) as the most
  promising production site.  If this is the
  case, the chemical enrichment timescale of the r- and s- processes
  are distinct, namely: a few tens of million years for the r-process
  nuclei (typical lifetime for a massive star), and more than 0.5 Gyr
  for the bulk of production of s-process in AGB stars
  \citep{Kaeppeler}.  Some production of s-process is also expected to
  take place in massive stars, although the overall contribution is
  negligible, especially in the very metal-poor regime
  \citep[see][]{Raiteri92}. For this reason, it has been common to
  associate the n-capture elements in the extremely metal-poor (EMP)
  stars of the halo to r-process production, as first pointed out by
  \citet{Truran81}.

  The solar system isotopic abundances \citep[see][]{Grevesse10} can
  be split into r- and s-process contributions.  Theoretically the
  s-process abundance pattern is known, and the solar s-process
  pattern is obtained by scaling to the abundances of isotopes
  exclusively produced by s-process (isotopes shielded against
  r-process by the corresponding stable isobar of Z$-$2).  The r-process
  contribution to the solar system \citep{Arla99,SSC04} is determined
  by subtracting the s-process contribution. In this way the so-called
  r-process contribution to the solar system is found, although is not
  free of theoretical uncertainties in the s-process contribution,
  chemical evolution, or other processes that could be hidden.

Although the relative amount of neutron capture elements is small
compared to those of the lighter elements (roughly $10^{-6}$), their absorption
lines in the stellar spectra can be observed and their abundances
calculated even for the most metal-poor stars.  Ever since the work by \citet{Sneden96} \citep[and
also][]{Sneden03,Sneden08}, remarkable agreement has been found between the
abundance pattern from Ba up to Th measured in a very r-process-rich
EMP star (CS 22892$-$052) and the r-process contribution to the solar
system.  This appeared to confirm both the
theory of the s-process from one side and the idea that r-process was
the only producer of (heavy) neutron capture in the early universe.

On the other hand, the same investigators, \citet{Sneden96} \citep[and
also][]{Sneden03,Sneden08}, have shown that the pattern for light
neutron capture elements (from Sr to Ag) does not always match the
so-called r-process solar pattern and \citet{Wasserburg96} have invoked
two r-processes to explain radioactive isotopes anomalies in solar
meteorites. Moreover, the ratio of light to heavy neutron capture
elements in other EMP stars is not constant, as expected for two
species coming from the same process. The light elements are more
enriched in some EMP compared to the expectations of a pure r-process
pattern, and show an increasing scatter towards low metallicities, up
to [Fe/H] $\sim -$4 \citep{Honda04, Franc07}.

A new process was invoked to provide the additional contribution for
light neutron capture elements and, at the same time, possibly produce
the spread in the ratio of light to heavy neutron capture elements.
\citet{Trava04} call this process the lighter element primary process
(LEPP), without specifying a possible r- or s- origin (although
typically s-processes are not a primary process). Later, when
analyzing the pattern of this possible missing process,
\citet{Montes07} obtained two viable solutions, one with typical
neutron fluxes of the r-process and the other closer to s-process
values, both of them able to provide roughly the correct pattern.

For the first time, it was shown in the work by \citet{Pigna08}, who
assume a primary amount of $^{22}$Ne guided by rotating models, that
rotating massive stars could produce a significant amount of s process
at low Z.  More recently, \citet{Frisch12} have calculated full stellar
evolution models using an extended nuclear reaction network and show
that fast rotation can boost the production of s-process at low
metallicities.

Here our goal is to test the impact of these yields in the early
chemical enrichment of the Galaxy.  Clearly, the s-process in
fast-rotating massive stars will not be the only contributor to the Ba
and Sr in the early Universe, since most of it comes from the still
uncertain r-process. This means that in our models, we need to also
include a production via r-process.  Unfortunately, given the
uncertainties discussed before, only a few groups provide r-process
stellar yields: for neutrino wind models \citep{ArcoMonte11}
and for O-Ne-Mg core collapse models \citep{WJM11}.  However, as
recently shown by \citet{Hansen13} for the case of Sr, these
predictions can still vary by more than 3 dex, and although the models
quoted above do produce Sr, they do not produce Ba. Other processes,
more promising for Ba production, such as neutron star merging or
magneto-rotationally driven SN have not yet led to detailed stellar yield
 computations.

In the present work we follow an alternative way to deal with the
problem of the absence of  robust theoretical yield predictions 
for the
r-process, already adopted by other authors such as \citet{Trava99,Trava04},
\citet{Ishimaru99}, and \citet{Cesc06}:
we adopt \emph{empirical yields}, which are extracted from the
available observational constraints (see Section 4.2.1). It
turns out that the \emph{empirical yields} we estimate are within the
uncertainties of the present r-process models. This will enable us to
quantitatively estimate the impact of \emph{spinstars} on the early
enrichment of heavy elements, as well as to guide theoretical efforts
on r-process nucleosynthesis.

\section{Observational data}
\subsection{Metallicity distribution function}

The chemical evolution model assumptions on star formation history and
winds are such that the overall shape of the metallicity distribution
(MDF) of the halo stars is reproduced.  The MDF is strongly affected
by the selection criteria of each sample, and most studies have tried
to take this into account.  A comparison with the results by
\citet{Yong13}, who focus on the extremely metal-poor tail of the
MDF, shows good agreement with the theoretical predictions of
\citet{Cesc10}, which is essentially the same model as is adopted  here
\citep[see Fig.3 of ][]{Yong13}.

Two other recent analyses making use of data from the Hamburg/ESO
objective-prism survey (HES) \citep{Christlieb08} have been recently
carried out with the goal of determining the whole halo MDF: the
\citet{Schoerck09} MDF based on metal-poor giants and the MDF from
\citet{Li10} based on main sequence turn off stars (MSTO) stars. Both
analyses use statically well understood selection criteria to obtain a
large number of metal-poor giants (1638) and metal-poor MSTO stars
(617), respectively.

Here we compare our model predictions with the MDF traced by MSTO stars rather than giants,
because as shown by  \citet{Li10}, the dwarfs are less affected by survey-volume corrections.
We are aware that this sample lacks the extremely metal-poor tail of stars with [Fe/H] $<-$3.5, but larger
statistically complete samples are required for this purpose. Fortunately, such
samples will be obtained from much larger and deeper surveys in the
near future, such as from SEGUE-2, LAMOST, SkyMapper, and 4MOST.


\subsection{Chemical abundances}

We have adopted observational abundance ratios from the literature;
the data for the neutron capture elements and for the
$\alpha$-elements are those collected by \citet{Frebel10}
\footnote{http://cdsarc.u-strasbg.fr/cgi-bin/qcat?J/AN/331/474},
labeled as halo stars\footnote{The list of the authors we use from the
  collection are \citet{MCW95}, \citet{MCW98}, \citet{ WES00}, \citet{
    AOK02b}, \citet{ COW02}, \citet{ IVA03}, \citet{ Honda04},
  \citet{AOK05}, \citet{ BAR05}, \citet{ AOK06b}, \citet{ IVA06},
  \citet{ MAS06}, \citet{ PRE06}, \citet{ AOK07c}, \citet{ Franc07},
  \citet{ LAI07}, \citet{COH08}, \citet{ LAI08}, \citet{ ROE08},
  \citet{ BON09}, \citet{HAY09}}. Among the halo stars collected, we
differentiate the normal stars from the carbon-enhanced metal-poor
(CEMP) stars.  Around 20\% of stars with [Fe/H] $< -$2.0 are CEMP
stars \citep{Lucatello06}.  We follow the definition given by
\citet{Masseron10} where a CEMP star is defined as having
[C/Fe]$>$0.9. The latter can be subclassified as
\cite[see][]{Masseron10}, based on Ba and Eu, namely:
\begin{itemize}
\item CEMP-s: [Ba/Fe]$>$1 and [Ba/Eu] $>$0 
\item CEMP-rs  [Eu/Fe]$>$1 and [Ba/Eu] $>$0 
\item CEMP-no [Ba/Fe]$<$1 and no Eu value
\item CEMP-r   [Ba/Fe]$<$1 and [Ba/Eu] $<$0
\item CEMP-low s [Ba/Fe]$<$1 and [Ba/Eu] $>$0.
\end{itemize}

We note that in our sample we have just one CEMP with [Ba/Fe]$<$1 with
Eu measurement, which is classified as CEMP-r , and so the CEMP-low s
category is missing.  For our purposes, the above categories are not
strictly necessary, but we prefer to refer to the literature than
create ``ad hoc'' definitions.  The important distinction in the
present work is between CEMP-(r)s and CEMP-no, hence whether a strong
signature of s-process is present or not.

Indeed CEMP-s most likely stem from binary mass transfer from a
previous AGB companion \citep{Bisterzo12,Lugaro12}, and for this
reason CEMP-s not reflect the chemical evolution of the ISM.

\section{The chemical evolution models}

Homogeneous chemical evolution models alone cannot take advantage of
the precious information recorded in the scatter of key element
abundance ratio of old stars, in different environments. The
inhomogenous chemical evolution models (i.e. models that relax the
instantaneous mixing approximation), in turn, requires a cosmological
framework in which it is possible to predict the properties (volumes)
of the inhomogeneities arising from both merger building blocks and
unmixed ISM (e.g., feedback). However, in this case other uncertainties
related to the field of cosmological simulations will play a crucial
role in the results.  Although we have started to work in this
direction, we still think that it is important to develop simpler
inhomogenous models, of the kind presented here.

Indeed, these models will serve as a test bench to study the different
nucleosynthetic prescriptions proposed by the different scenarios.
The goal is to identify key abundance ratios, as well as the most
promising stellar yields, to be implemented in cosmological
simulations. This step is crucial for avoiding spending unnecessary CPU
time with models that do not meet the minimum requirement set by a
first comparison with observations.

The neutron capture element spread in EMP stars has already been the
subject of other investigations.  We recall here the works of
\citet{Tsujimoto99}, \citet{Ishimaru99}, \citet{Trava01},
\citet{Argast02,Argast04}, and \citet{KaGu05}.   Most of these models
reproduced the spread in n-capture elements observed in EMP stars, but
left unexplained other problems: a much smaller scatter observed for
the other elements \citep[for instance $\alpha$-elements,
see][]{Cayrel04,Boni12}, the reduction of the spread toward higher
metallicity for the neutron capture elements, and the large scatter
observed in N/O and C/O.

This has pointed to a different solution in which the inhomogeneities
arise from a stochastic formation of massive stars.  The spread can be
generated by the enrichment of different species if they are produced
by different ranges of masses, as shown for heavy neutron capture
elements as a function of iron in \citet{Cesc08};
at the same time, the spread is reduced after a
few generations of stars, and it is small for elements that share
the same range of production (as $\alpha$-elements and iron). This
approach has also been used for the CNO elements \citep[see
][]{Cesc10} to investigate the impact of the inhomogenous modeling on
the observed spread (in particular, of the ratio N/O) in EMP stars.

Here we consider the same chemical evolution model as adopted in
\citet{Cesc10}, based on the inhomogenous model developed by
\citet{Cesc08} and on the homogeneous model of \cite{Chiappini08}. We
consider that the halo consists of many independent regions, each with
the same typical volume, and each region does not interact with the
others.  We decided to have a typical volume with a radius of roughly
90 pc, and the number of assumed volumes is 100 to ensure good
statistical results. The dimension of the volume is large enough to
allow us to neglect the interactions among different volumes, at least
as a first approximation.  In fact, for typical ISM densities, a
supernova remnant becomes indistinguishable from the ISM -- i.e.,
merges with the ISM -- before reaching $\sim50pc$ \citep{Thornton98},
less than the size of our region. We do not use larger volumes because
we would lose the stochasticity we are looking for; in fact, larger
volumes produce more homogeneous results.

In each region, we assume the same law for the infall of the gas with
primordial composition, following the homogeneous model by
\citet{Chiappini08}, so we have the following Gaussian law:
\begin{equation}
\frac{dGas_{in}(t)}{dt} \propto e^{-(t-t_{o})^{2}/\sigma_{o}^{2}} 
\end{equation}
where $t_{o}$ is set to 100Myr and $\sigma_{o}$ is 50Myr.
Similarly, the star formation rate (SFR) is defined as
\begin{equation}
SFR(t) \propto ({\rho_{gas}(t)})^{1.5}
\end{equation}
where $\rho_{gas}(t)$ is the density of the gas mass inside the considered volume.  
Moreover, the model takes an outflow
from the system into account:
 \begin{equation}
\frac{dGas_{out}(t)}{dt}  \propto SFR(t).
\end{equation}
\begin{figure}
\begin{center}
\includegraphics[width=.49\textwidth]{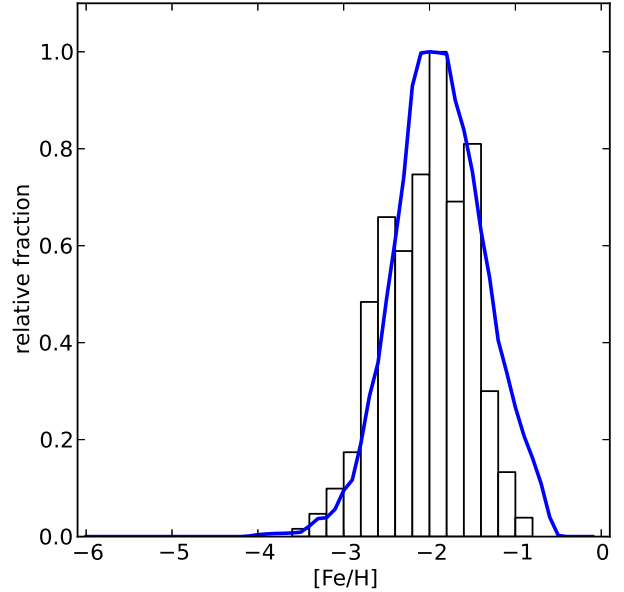}
\caption{Comparison between the model MDF in blue line and the observed halo MDF by Li et al. (2010):
main-sequence turnoff stars in the HESS (Hamburg ESO) for which [Fe/H]$<-1$.}\label{fig0}
\end{center}
\end{figure}
\begin{figure*}[ht!]
\begin{minipage}{190mm}
\begin{center}
\includegraphics[width=160mm]{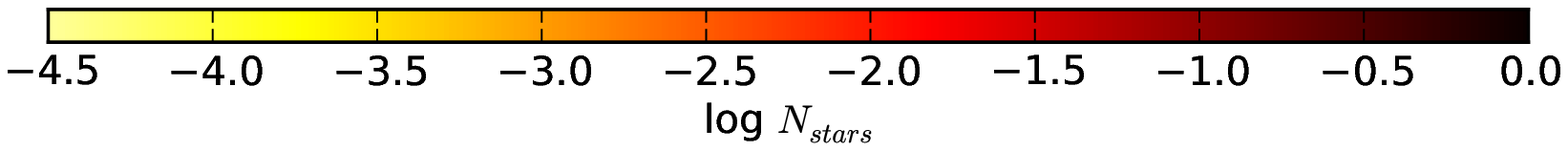}
\includegraphics[width=160mm]{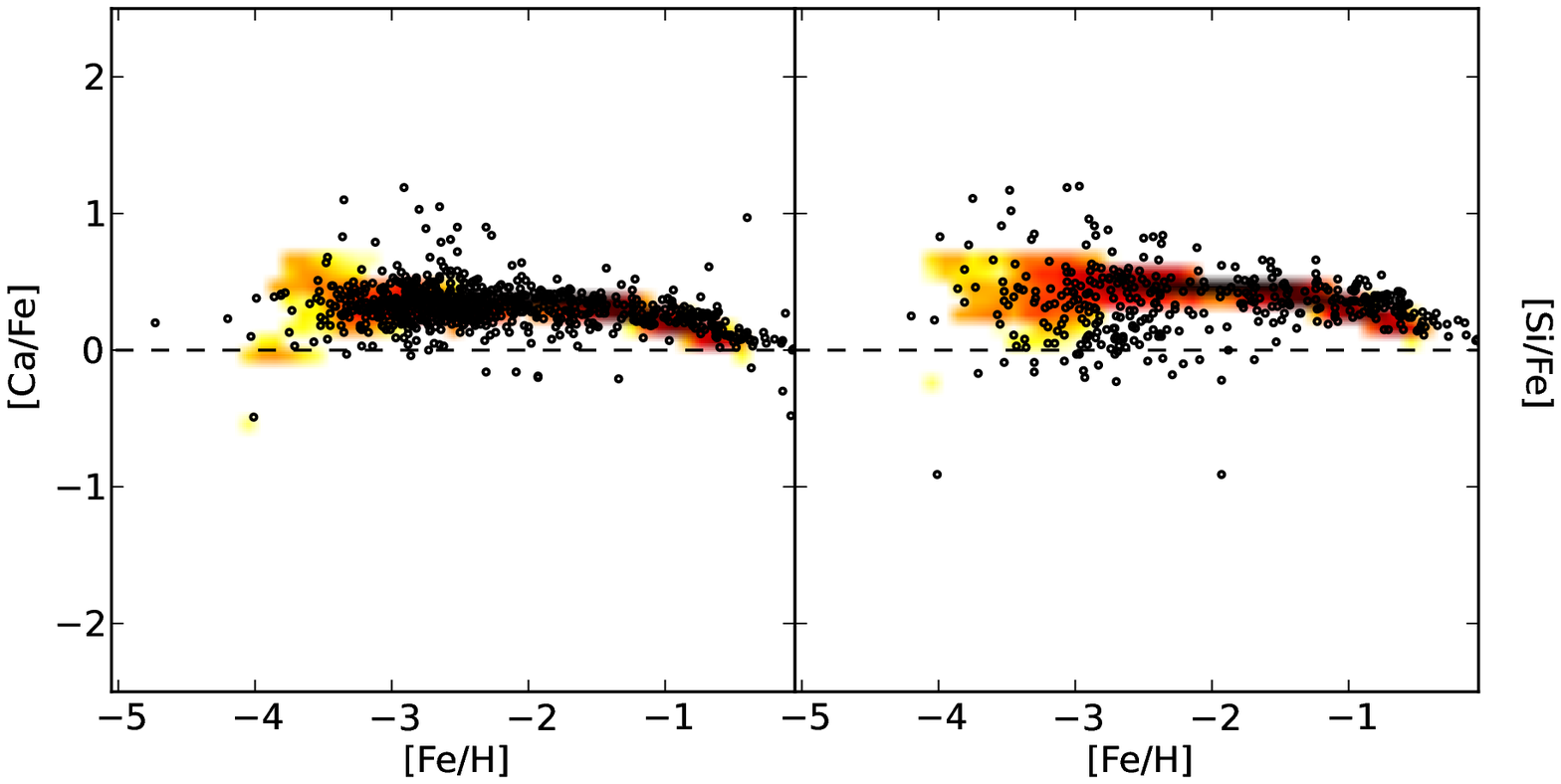}
\caption{[Ca/Fe] and [Si/Fe] vs [Fe/H], from left to right.  The
  density plot is the distribution of simulated long-living stars for
  our model; the density is on a logarithmic scale, normalized to the
  peak of the distribution and the bar over the plot describes the
  assumed color scale.  Superimposed on the density plot, we show the
  abundances ratios for halo stars  \citep[data from][]{Frebel10}.}\label{fig1b}
\end{center}
\end{minipage}
\end{figure*}
Knowing the mass that is transformed into stars in a timestep
(hereafter, $M_{stars}^{new}$), we therefore assign the mass to one
star with a random function, weighted according to the initial mass
function (IMF) of \citet{Scalo86} in the range between 0.1 and
100$M_{\odot}$.  Then we extract the mass of another star, and we
repeat this cycle until the total mass of newly formed stars exceeds
$M_{stars}^{new}$.  In this way, in each region at each timestep, the
$M_{stars}^{new}$ is the same, but the total number and mass
distribution of the stars are different. We then know the mass of each
star contained in each region, when it is born, and when it will die,
assuming the stellar lifetimes of \citet{MM89}.  At the end of its
lifetime, each star enriches the ISM taking the enrichment
due to that star into account and owing to the already present mass of each element
locked in that star when it was born.

Yields for $\alpha$-elements and Fe are the same as in
\citet{Franc04}.  The model considers the production by
s-process from 1.5 to 3M$\odot$ stars and SNIa enrichment, as in
\citet{Cesc06}.  The model does not include the most recent and
  accurate yields for AGB, and the very low-metallicity yields for AGB
  are extrapolated.  However, due to the longer lifetime compared to
the halo star formation history, this production is hardly seen in
the results of the model for [Fe/H] $<-$1, as well as in the abundance
pattern of normal EMP.   Super-AGB stars can also  be involved in
  the production of neutron capture elements, but at the present time
  yields for those elements from SAGB are not available; 
  however, the production of s-process elements in their convective
  pulses is expected to be small \citep{Siess10}.  Additional studies
  are required to clarify this aspect. 

\subsection{First test: reproducing the MDF and the [$\alpha$/Fe] low scatter}

This model is able to reproduce the MDF measured for the halo
by \citet{Li10} based on HES data of main sequence turn-off stars 
(see Fig. \ref{fig0}). This comparison shows that the timescale of
enrichment of the model looks like that of the halo stars in the
solar vicinity.

In Fig. \ref{fig1b}, we show the predictions of our model for the
$\alpha$-elements Ca and Si.  Our model predicts a narrow spread for
these abundance ratios, which is compatible with the observational data.  In
addition, our model predictions show a slightly increase in the
scatter in the very-metal-poor regime.  Interestingly, inhomogenous
models, as the one shown here, do predict a few
outliers with low values of [$\alpha$/Fe].

\subsection{Stellar yields for heavy elements}
\label{sec:yields}

\subsubsection{\emph{Empirical} yields for the r-process}

As mentioned in Section~\ref{sec:heavy}, the site of production of
r-process elements is still a matter of debate.  In particular, most
of the neutron reactions involving r-process elements are beyond the reach
of nuclear physics experiment. In such a situation, we must use
observational data to guide the theoretical models for this process.

\begin{figure*}[ht!]
\begin{center}
\includegraphics[width=.49\textwidth]{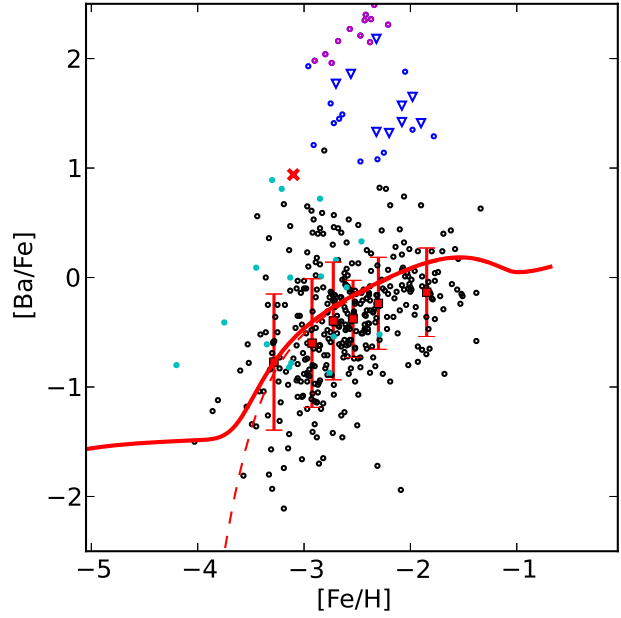}
\caption{[Ba/Fe] vs [Fe/H] abundances ratios of the halo stars
\citep[data from][]{Frebel10}: 
black open circle are normal stars, 
blue open circle  for CEMP-s stars (open triangles without Eu value)
magenta open circle for CEMP-rs,
cyan filled circle for CEMP-no,  and 
red x marker for the CEMP-r star.
 The  error bars represent the mean and the standard deviations for the
  normal stars abundances calculated over different bins in [Fe/H].
  The bins are calculated in such a way that each bin contains the same
  number of data points. The results of the homogenous model with the assumed
  \emph{empirical} yields for Ba is shown by the solid line; the dashed line shows the results
when only the ``standard'' r-process site is considered. }\label{fig2}
\end{center}
\end{figure*}

\begin{figure*}[ht!]
\begin{minipage}{185mm}
\begin{center}
\includegraphics[width=180mm]{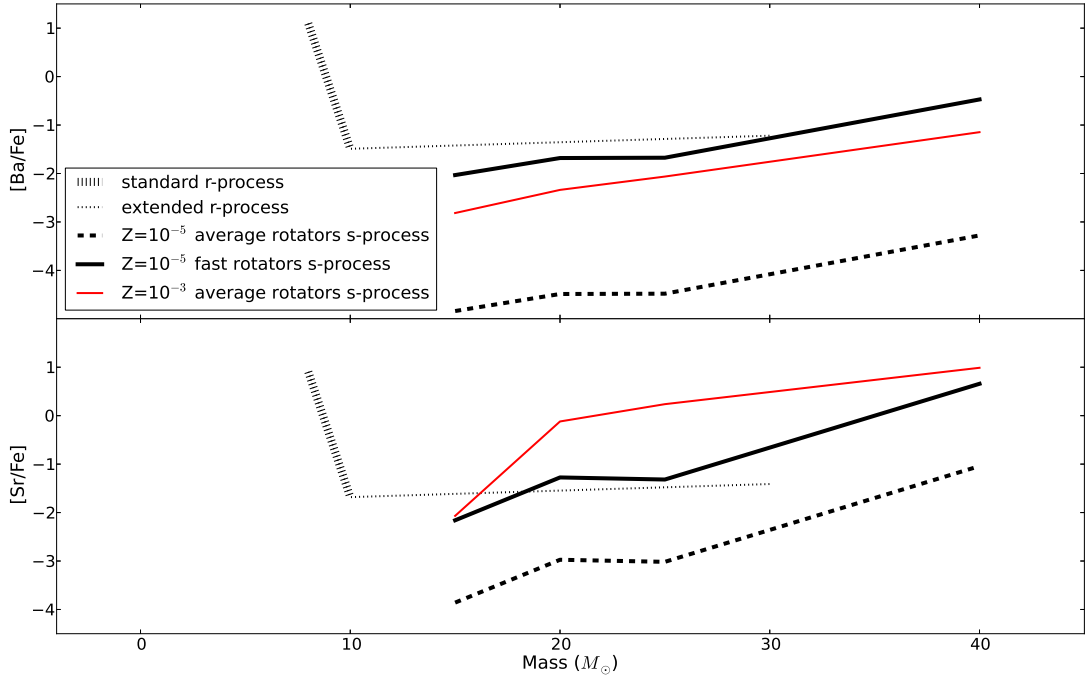}
\caption{
Yields used, normalized to the solar (Grevesse et al. 2010), 
for the ratios of [Ba/Fe] and [Sr/Fe], as a function of the stellar mass
and the metallicities}\label{fig3}
\end{center}
\end{minipage}
\end{figure*}

Given the above, we adopt here the following approach \citep[see
also][]{Cesc06}: we compute a homogeneous chemical evolution model
where the yields of Ba are chosen so as to reproduce the mean trend
of [Ba/Fe] versus [Fe/H] (see Fig.~\ref{fig2}).  The resulting yields
are what we call \emph{empirical} yields here.

The \emph{empirical} yields are obtained as the simplest array able to
reproduce the observed trend of increasing [Ba/Fe] with metallicity,
but also present one important property, namely, the need for two
different sites of production (see Fig. \ref{fig3}):
\begin{itemize}
\item a strong production in a narrow mass range 8-10$M_{\odot}$
that we call \emph{standard} r-process site;
\item an extended contribution coming from stars with higher masses
  (from 10 up to 30$M_{\odot}$), whose contribution is lower by $\sim$
  2 dex than that of the \emph{standard} r-process, and which we here
  denote as the \emph{extended} r-process site.
\end{itemize}

 The \emph{standard} r-process assumptions fit the general
  concept of r-process production, so an intense production in
  relatively rare events. On the other hand, the \emph{extended}
  r-process seems a necessary step to explain the presence of stars
  with [Ba/Fe] $\sim -$0.7 at metallicities lower than [Fe/H] $\sim
  -$3.5 (Francois et al. 2007, see Fig. \ref{fig3}). This effect can
  be appreciated better from the inhomogenous results (see
    comparison of model results with and without the \emph{extended}
    r-process in Fig. \ref{appendix} in the appendix), and it is a viable
  way to explain the absence of stars lacking Sr and Ba
  \citep{Roederer13}, as in the case without the \emph{extended}
  r-process.

  We here assume the r-process yields of Sr scale with those of Ba.
  The adopted scaling factor (of 3.16) has been taken from the solar
  system r-process contribution (see Sect \ref{sec:heavy}) as
  determined by \citet{SSC04}.  Another meaningful assumption could
  have been the choice of the mean observed ratio between Sr and Ba in
  r-process-rich stars (see Cowan et al. 2006). In this case, the Sr/Ba
  ratio would be around 0.2 dex higher. On the other hand, this same
  ratio measured in the r-process rich star CS22892$-$052 is close to
  the value that we have adopted.  We stress that we have 
    selected a simple approach here to the r-process production, although
    many authors have recently started to suspect that this assumption
    may be not completely valid \citep{Roederer10,Boyd12}; on the
    other hand, our simple approach is reasonable, since the emphasis
    of this paper is not on the r-process issue but on the
    contribution from spinstars to the production of strontium and
    barium.

Interestingly the stellar mass range for \emph{standard} r-process is
close to the one predicted by theoretical models of ONeMg core
supernovae (with initial masses toward the lower end of the massive
stars). However, the latest models \citep{WJM11, WNJ09} do
not succeed in producing heavy neutron capture elements such as Ba.
 It is also worth mentioning that the stellar mass range 8$<
  M_{\odot} <$ 10 is in the transition between the super AGB stars
  and the electron capture supernova. For this reason we stress that
  the final fate of these stars is still rather uncertain \citep{Siess07, Siess10}. 

  As we see, the role of the \emph{extended} r-process contribution,
  which we need to consider when empirically explaining the observed
  [Ba/Fe] and [Sr/Fe] in EMP halo stars, can be replaced by the
  contribution of \emph{spinstars}.
 
\subsubsection{The contribution of \emph{spinstars}}

To illustrate the impact of rotating stellar models on the chemical enrichment 
of Sr and Ba in the earliest phases of the Universe, we now focus on three sets 
of inhomogeneous chemical evolution models, computed with the following 
sets of stellar yields (see Fig.~\ref{fig3} and Table 1):
\begin{itemize}

\item {\bf r-model}: assume only \emph{empirical yields} for the r-process (\emph{standard} + \emph{extended});
\item {\bf as-model}: assume only the \emph{standard}  r-process yields (no \emph{extended} r-process contribution),
 plus the s-process yields coming from rotating stellar models (see below) 
\item {\bf fs-model}: similar to the model above, but with s-process
  yields coming from the s-process in fast-rotating stellar
  (spinstars) models.
\end{itemize}

The nucleosynthesis adopted in the {\bf as-model} for the s-process
comes from unpublished results by Frischknecht (2011, PhD thesis).  In
this set of yields, the s-process for massive stars is computed for
$v_{ini}/v_{crit}$=0.4 and for a standard choice for the reaction
$^{17}O(\alpha,\gamma)$ from \citet{CF88}; they are composed of a grid
of four stellar masses (15, 20, 25 and 40$M_{\odot}$) and three metallicities
(solar metallicity, $10^{-3}$, $10^{-5}$); in Fig.~\ref{fig3} we show
the yields for the two lowest metallicity cases.  We do not extrapolate
the production toward stars that are more massive than 40$M_{\odot}$ (although
it is realistic also to have a production in this range), but we
extrapolated the $Z=10^{-5}$ grid down to Z=0.  In addition to the
s-process in massive stars, we take our empirical
r-process enrichment into account but coming only from the \emph{standard}
r-process site. In this way we are decoupling the sites of production
for the two processes. Interestingly, this figure  shows that the
  s-process of \emph{spinstars}  plays the role here of the \emph{extended}
  r-process, which is not used in this model.

It is worth pointing out that the yields we have used for the as-model
are very conservative, among the models computed by \citet{Frisch12}
for 25$M_{\odot}$.  In the {\bf fs-model}, we have adopted a more
robust production of s-process, which is achievable if we consider the
s-process yields of massive stars computed for $v_{ini}/v_{crit}$=0.5
(fast rotators) and for a reaction rate $^{17}O(\alpha,\gamma)^{21}Ne$ one
tenth of the standard choice. Indeed, the adopted $v_{ini}/v_{crit}$=0.4 in
the as-model is lower than the value we have used in previous studies
of spinstars (0.6-0.8) \citep{Chiappini06, Chiappini08}. In addition,
recent star formation simulations \citep{Stacy11,Stacy12} support high
rotation velocities compatible with higher $v_{ini}/v_{crit}$ ratios.
Furthermore, the current uncertainties in the $^{17}O(\alpha,\gamma)$
rate are still large. The adopted value of one tenth of the standard
one is well within both the theoretical and recent experiments
uncertainties.

We do not have a fully computed grid for these parameters, but we did
scale the previous yields guided by the results obtained with these
parameters for a 25$M_{\odot}$ of $Z=10^{-5}$ by \citet{Frisch12}
(cfr. in their paper Table 2), and applied the same factor to all the
masses at $Z=10^{-5}$, as shown in Fig. \ref{fig3}. Again, for this
{\bf fs-model}, in addition to the s-process in \emph{spinstars}, we
took the contribution by our empirical \emph{standard}
r-process into account.

 Finally we note that \citet{Frisch12} only provide the
  pre-supernovae yields. However, the supernova shock only affects the
  bottom of the carbon shell and not the whole s-process-rich
  region. The s-process takes place at the end of core
    He-burning, and the start of shell carbon burning. At low Z, as
    explained in \citet{Frisch12}, the contribution from carbon
    shell burning is very small because of the strong neutron poisons
    during this phase. The composition of the s-process-rich layers
    are thus set by the end of He-burning, and very little happens
    during the advanced stages at very low Z.  As a consequence the
  bulk of the s-process-rich material remains almost unaffected by the
  SN explosion \citep[also the case of light elements such as C, N,
  and O - see ][]{Woosley02}. A more recent study by \citet{Tur09}
  shows that certain isotopes can be affected by both explosive
  nucleosynthesis and also shell mergers possibly taking place during
  the early collapse. These changes will, however, not affect
  elemental ratios like [Sr/Ba] and the conclusions of our paper.

\begin{table*}
\begin{center}
\caption{Nucleosynthesis prescriptions for the three cases.
}
\begin{tabular}{|c|c|c|c|}
 \hline
  Model name& panels in Fig. \ref{fig4}    &   s-process    &  r-process      \\  
\hline
 r- & Upper        & No s-process from massive stars & standard + extended r-process site (8 - 30 M$_{\odot}$)\\
 \hline
 as- & middle    &    average rotators ($v_{ini}/v_{critic}=0.4$)& standard r-process site  (8 - 10 M$_{\odot}$)\\
\hline
 fs- &lower    &   fast rotators ($v_{ini}/v_{critic}=0.5$)      & standard r-process site (8 - 10 M$_{\odot}$) \\
 && and 1/10 for $^{17}O(\alpha,\gamma)$ reaction rate & \\
\hline
\hline 
 \end{tabular}
\end{center}
\
\label{tabmodels}
\end{table*}

\begin{figure*}[ht!]
\begin{minipage}{185mm}
\begin{center}
\includegraphics[width=185mm]{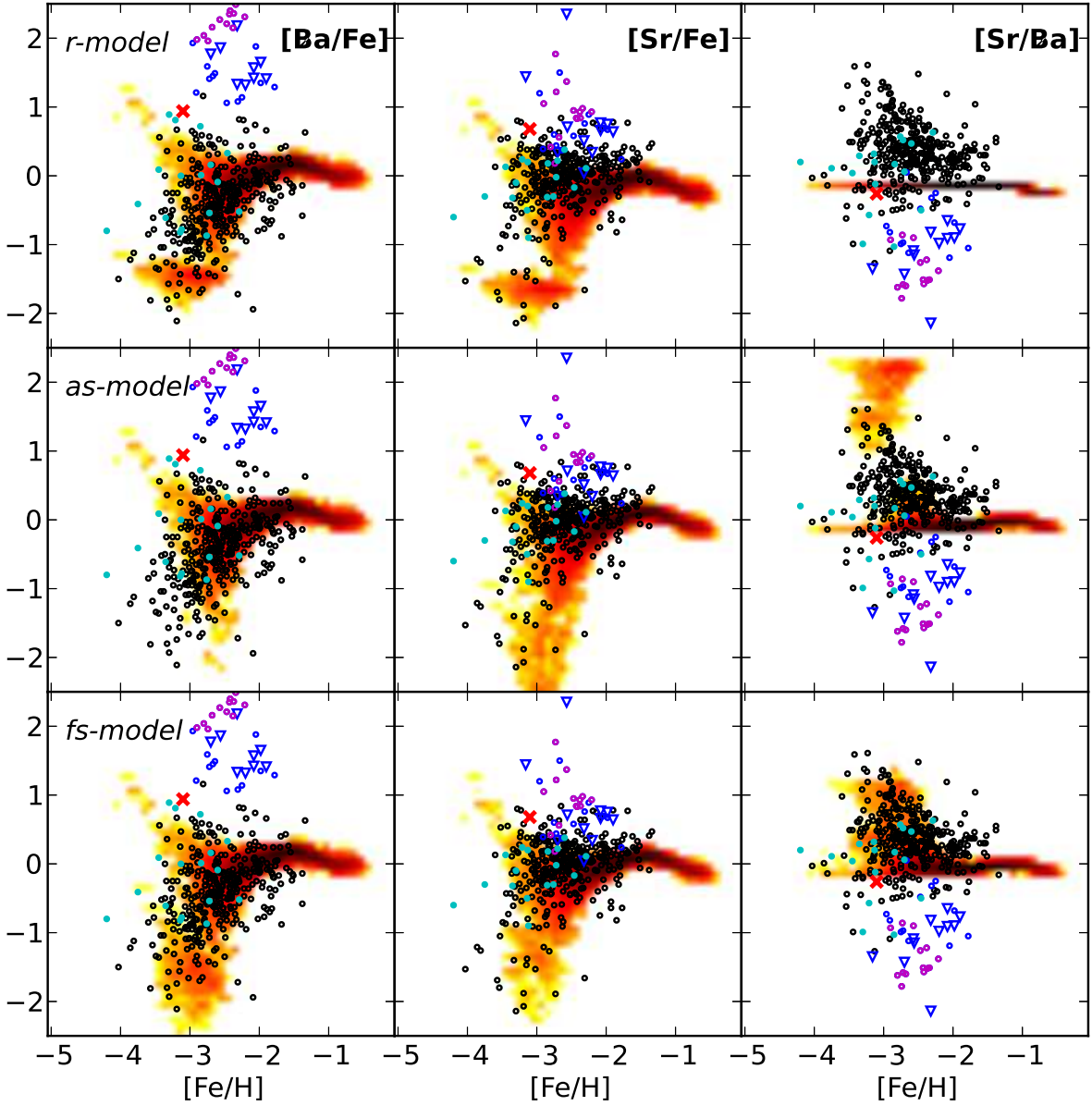}

\caption{On the left [Ba/Fe], in the center [Sr/Fe], on the right [Sr/Ba],
 vs [Fe/H]. Upper, center, and lower panels for  ``r-'',''as-", and "fs-"models respectively. 
The density plot is the distribution of simulated  long-living stars for our models;
the density is on a log scale, normalized to the peak of the distribution
(see bar over the Fig. ~\ref{fig1b} for the color scale).
Superimposed, we show the abundances ratios for halo stars \citep[data from][]{Frebel10}.
The symbols are the same as in Fig. ~\ref{fig2}.
}\label{fig4}

\end{center}
\end{minipage}
\end{figure*}

\section{Results}

Our main results are summarized In Fig.\ref{fig4}.  We started by
analyzing the results of the {\bf r-model} (upper panel), which
only assumes the contribution from massive stars via our
\emph{empirical} r-process yields.  The spread obtained by this model
matches the dispersion of [Ba/Fe] (upper, left panel), confirming that
the hypothesis of a contribution by two distinct sites seems to be a
good one to explain the data. In the particular case of this model,
the two sites of production are illustrated by a \emph{standard}
r-process that takes place in the lower mass range of the massive
stars, and an \emph{extended} r-process taking place in more massive
stars. The first one is assumed to be much more efficient (larger
quantities of ejected material) than the second.

  We also consider the match with the observed [Sr/Fe] obtained by the same
model to be a good match, especially at very 
low metallicities, because more than 80\% of the stars below [Fe/H]$<-$3 are
within the model results.  Since the Sr yields
adopted here are obtained just by using the Ba/Sr ratio matching the
observed solar system r-process \citep{Sneden08}, the results for this ratio are simply
constant with metallicity (as shown in the upper right panel). This
suggests that some other physical process, taking place in the
same mass range as what we call \emph{extended} r-process might
be contributing to producing part of the Sr and Ba in the very early
Universe (since the scatter is larger between metallicities $-$2.5 and
$-$3.5). In other words, the {\bf r-model} is useful for highlighting what
we are trying to solve in the present work.

We now turn to the results obtained with our {\bf as-model} (see Table
1 and Fig. \ref{fig4}, second row), and see the impact of the
s-process production of Ba and Sr by rotating massive stars on the
halo chemical evolution. In this case we added a rather conservative
s-process production from rotating stars, and turned off the
contribution of what we have called \emph{extended} r-process.  The
results for [Ba/Fe] and [Sr/Fe] are not completely satisfactory
because the model cannot reproduce the low [Ba/Fe] ratios observed in
extremely metal-poor stars with [Fe/H]$<-$3 (left and middle panels of
the second row). This happens because this rather conservative model
does not predict enough Sr and, particularly, Ba (this can be seen in
Fig.~\ref{fig3} by comparing the stellar yields of the \emph{extended}
r-process -- dotted line -- with the yields of the {\bf as-model} at
$Z=10^{-5}$ -- dashed line).

Despite the shortcomings described above, the interesting result of
this model resides in the [Sr/Ba] plot (second row, right panel),
where it is clear that this new process produces an overabundance and
creates a spread in [Sr/Ba] at [Fe/H] $\sim -$3.  From this, one can
conclude that the s-process produced in fast-rotating massive stars
seems to act in the correct direction.  Nevertheless, these results
are affected by the ratio of the yields of Sr and Ba predicted by this
specific stellar evolution model \citep[similar to model B1 of
][]{Frisch12}. These prescriptions produce [Sr/Ba]$>$2, whereas the
EMP stars show  [Sr/Ba]$\sim$1.5 at the maximum.

Finally for the {\bf fs-model} (Fig. \ref{fig4}, bottom panels), where
we adopted less conservative stellar models predicting even greater
s-process enrichment (essentially due to a higher $v_{ini}/v_{crit}$
and a lower $^{17}O(\alpha,\gamma)$ reaction rate similar to the model
B4 of Frischknecht et al. 2012 -- see Section~\ref{sec:yields}), the
theoretical predictions and observations show striking agreement.
This model not only reproduces the [Ba/Fe] and [Sr/Fe] scatter closely
(left and middle bottom panels), but can also account for the observed
[Sr/Ba] spread (right bottom panel), at the correct metallicity
interval.
This shows that with a less conservative production of s-process in
fast-rotating massive stars (as is the case in the {\bf fs-model}),
this process could play the same role as our \emph{extended} r-process
in the {\bf r-model}. In addition, it shows that a \emph{standard}
r-process (taking place in the lower mass range of the massive stars)
enters into play with a weight that increases as the metallicity
increases (still in the very metal-poor range).

In the scenario of the fs-model, the stars with high [Sr/Ba]
abundances should show the pollution by the s-process in
\emph{spinstars}. These stars should therefore present a high value of
the even/odd isotopes of Ba (as typical of an s-process).
Interestingly, the pioneering measurements of Ba isotopes  in six EMP
stars by \citet{Gallagher10,Gallagher12} have shown exactly the
feature we predict with the above models, namely, a high level of even
isotopes compatible with an s-process production. It is
  important to note that four of these stars, with Sr abundances available
  in literature, present high [Sr/Ba] ratios, excluding AGB
  pollution; for the remaning two stars, a similar conclusion is
  supported by the Y abundance. We are aware that even with the
highest resolution and signal-to-noise available, the measurement of
the ratio of odd-to-even isotopes of Ba could still suffer from large
uncertainties (such as broadening of the studied lines due to the
differential hyperfine splitting of its isotopes). More measurements
would be needed to confirm our suggestions. We stress, however, that
at present, only our scenario can explain these observations.

\begin{figure}[h!]
\begin{center}
\includegraphics[width=.49\textwidth]{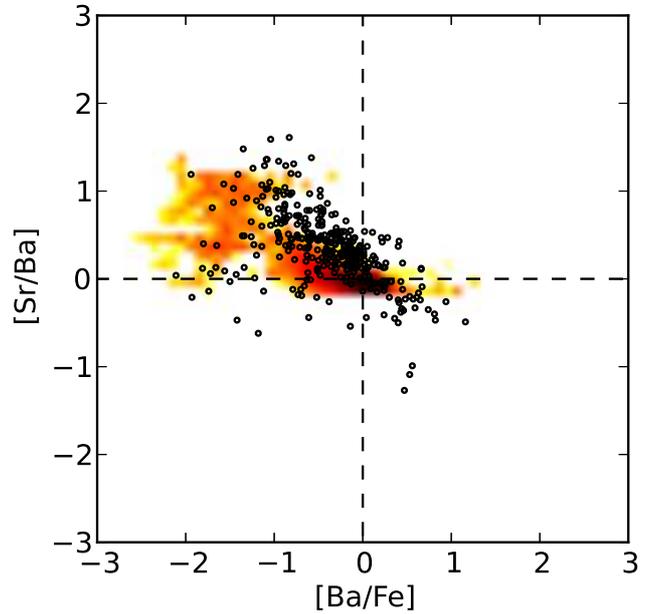}

\caption{[Sr/Ba] vs [Ba/Fe], the density plot is the distribution of
  simulated long-living stars for fs-model. The density is on a log
  scale, normalized to the peak of the distribution (see bar over the
  Fig. ~\ref{fig1b} for the color scale).  Superimposed on the density
  plot, we show the abundances ratios only for normal halo stars
  \citep[data from][]{Frebel10}.}\label{fig5}

\end{center}
\end{figure}

In Fig. \ref{fig5}, we plot the [Sr/Ba] vs [Ba/Fe] similarly to the
plots in the works by \citet{Montes07} and \citet{Franc07}.  Our model
can reproduce the peculiar behavior seen in EMP stars, namely, a high ratio
of [Sr/Ba] at low [Ba/Fe] and the correct amount of scatter.
This behavior in the model is determined by the fact that by including
the s-process production in \emph{spinstars}, we are able to decouple
the sites responsible for the r-process process
enrichment (8-10M$_{\odot}$), from the producers of s-process in the
early Universe (i.e. the most massive \emph{spinstars}, which are
responsible for the largest contributions). In volumes where the most
massive stars contribute more, the production of s-process will be
important, and the [Sr/Ba] ratios will be high. However, in this case
the net enrichment in Ba and Sr will be low (due to the low yields
produced by this process), which will then lead to low [Ba/Fe] and
[Sr/Fe] ratios. On the other hand, in volumes where the contribution
of the less massive stars is greater, the standard r-process will tend
to produce higher levels of [Ba/Fe] and [Sr/Fe] enrichment, whereas
the [Ba/Sr] will tend to be lower. The different mixtures of these two
limits is dictated by the stochasticity of the star formation rate of
each independent volume, which then lead to the observed spread.

\section{Discussions and conclusions}

We have developed an inhomogeneous chemical evolution model for the
halo with the aim of explaining the observed scatter (or lack of) in
the abundance ratios of key chemical elements in very metal-poor
stars. The models presented here serve as a test-bench to study the
different nucleosynthetic prescriptions proposed by different stellar
evolution groups.  The goal is to identify the most promising stellar
yields for key abundance ratios (in this case neutron capture
elements) to then be implemented in cosmological simulations (a
project that we are already pursuing).  This step is crucial to avoid
spending unnecessary CPU time with models that do not meet the minimum
requirements set by a first comparison with observations. In addition,
the comparison of our model predictions with observations is very
useful for offering additional constraints to the theoretical
nucleosynthesis predictions, which are still very uncertain for these
elements.

Our main results and conclusions are summarized below.

\begin{itemize}

\item We first showed that it is possible to reproduce the observed
  spread of [Ba/Fe] and [Sr/Fe] and the simultaneous lack of scatter
  in the $\alpha$-elements in EMP stars with an inhomogeneous
  chemical evolution model, if we assume a production of neutron
  capture elements coming from: a) a relatively rare, but efficient,
  site of production (here illustrated by a contribution of stars in
  the 8-10 M$_{\odot}$ mass range), and b) a second site of
  production, which is less efficient when producing lower amounts of
  neutron capture (here illustrated with a contribution from stars in
  the 10-40 M$_{\odot}$ mass range). More important, we show that
  despite the good agreement with the observed spread of [Ba/Fe] and
  [Sr/Fe], an extra process is needed to explain the observed [Sr/Ba]
  scatter.
\item The presence of r-process-rich stars, with a common strong
  r-process signature, is an observational constraint for the first
  site of production: only if this site produces strong enhancement in
  relatively rare events can the chemical evolution model then  
   provide an explanation for the r-process-rich stars. Here we show
  that keeping this first site of production fixed, but adding the
  contribution to the neutron capture elements by fast-rotating
  massive stars (now considered as the second site of production), then it
  is possible to create scatter in the ratio of [Sr/Ba], as
  observed.

\item In particular, when considering stellar models for fast-rotating
  massive stars that are less conservative in their assumptions (what
  we call here spinstar s-process models), we are then able, for the
  first time to our knowledge, to reproduce simultaneous the
  [$\alpha$/Fe], [Sr/Fe], [Ba/Fe], and [Sr/Ba] trends and scatter
  observed in halo stars with an inhomogenous chemical evolution
  model.  Although the proposition that the scatter in the observed
  spread between heavy and light neutron capture elements in EMP stars
  could be due to the contribution of \emph{spinstars} had been made
  before \citep{Chiappini11}, but here we show quantitative estimates
  that seem to confirm this hypothesis. In this elegant
  solution, we are able to explain the results with just two different
  sites of production for Sr and Ba, without requiring more
  complicated scenarios.

\item The above solution is based on the existence of \emph{spinstars}
  in the early Universe.  Indeed, this is not the first chemical
  signature to reveal the importance of fast-rotating massive stars in
  the early chemical enrichment of the Galaxy. On the contrary, very
  old halo stars were previously found to show at least four other
  abundance anomalies, which rotating models of massive stars can
  successfully account for:   rise in N/O and C/O \citep{Chiappini06}, low
  $^{12}$C/$^{13}$C \citep{Chiappini08} and a primary-like evolution
  of Be and B \citep{Prantzos12}. What we provide here is a fifth
  independent signature: an early enrichment of the Universe
  in s-process elements.

\item Our model thus predicts that stars with high [Sr/Ba] abundances
  should show the pollution by the s-process in
  \emph{spinstars}. These stars should therefore present a high value
  of the even/odd isotopes of Ba (typical of an s-process).  We are
  aware that even with the highest resolution and signal-to-noise
  available, the measurement of the ratio of odd-to-even isotopes of
  Ba could still suffer from large uncertainties, but measurements of
  this ratio would be needed to confirm our suggestions.

\item
Our results also show that to reconcile the model to the observations,
we need an r-process site of production decoupled from the one of
s-process. We assumed here the most straightforward solution, which is to
consider two different mass ranges for the two processes, but other
solutions providing a large amount of r-process production in only a
fraction (roughly 15\%) of the massive stars would be equally valid, as in the
case of magnetorotationally driven supernovae \citep{Winteler12}.

\item
At variance with suggestions so far in the literature of a possible second r-process or
weak r-process to explain the observations, the production of s-process in fast-rotating
massive stars, does predict an enrichment in Ba (although not as
strong as for Sr), and a negligible production of Eu. Another
natural outcome of our model is thus the production of a small spread ($\sim$0.3 dex)
in the ratio of [Eu/Ba], which a scenario involving two different r-processes cannot 
reproduce. The [Eu/Ba] spread is observed in stars with [Fe/H]$\sim -$3, although more data is needed 
to probe this additional, valuable observational constraint.

\end{itemize}

 We are aware that the recent simulations of star formation in the
  early universe find that most of the first stars are formed in
  binary systems \citep{Stacy12}.  However, binary stars, just like
  single stars, should also experience rotation-induced mixing that is
  the key property for producing of s-process in spinstars (as
  shown for the extreme case of long GRB progenitors by
  \citet{Cantiello07}. Indeed, the above simulations do predict the
  first stars to have high initial rotational velocities.

As a final comment we notice that, intriguingly, the CEMP-no stars
appear indistinguishable on the basis of the neutron capture elements
from normal EMP stars. In \citet{Cesc10}, our inhomogenous model with
fast-rotating massive stars was not able to account for the C/O and
N/O ratios in the same CEMP-no stars.  As mentioned in
\citet{Cesc10}, the explanation can reside in a pollution by a process
that produces C in a non standard chemical enrichment fashion
(ie. without mixing with all the surrounding ISM).  That no
strong s-process signature is observed in these stars could support
the scenario described in \citet{Meynet10}, in which a fast-rotator
star enriches low-mass stars belonging to its forming stellar cluster
through its stellar wind, highly enriching them in C, without
strong dilution by the ISM. In this case, no s-process can pollute
these low-mass stars because the s-process in fast-rotating massive
stars are not removed by wind, because not present on the surface.
However, we cannot exclude a scenario in which also CEMP-no are formed
by an AGB companion with a dimmed production of s-process
\citep[although most of the CEMP-no appear to be single stars at the
present time, see][]{Norris13}.

\begin{acknowledgements} 
  We thank the two referees, one theoretical and one observational,
  for the helpful suggestions and comments that have improved this
  manuscript.  R. Hirschi acknowledges support from the World Premier
  International Research Center Initiative (WPI Initiative), MEXT,
  Japan; R. Hirschi and C. Chiappini acknowledge support from the
  Eurogenesis EUROCORE program.  The research leading to these results
  has received funding from the European Research Council under the
  European Union's Seventh Framework Program (FP/2007-2013) / ERC
  Grant Agreement n. 306901.
\end{acknowledgements}

\bibliographystyle{aa}
\bibliography{spectro}

\begin{thebibliography}{130}
\expandafter\ifx\csname natexlab\endcsname\relax\def\natexlab#1{#1}\fi

\bibitem[{{Abel} {et~al.}(2002){Abel}, {Bryan}, \& {Norman}}]{Abel02}
{Abel}, T., {Bryan}, G.~L., \& {Norman}, M.~L. 2002, Science, 295, 93

\bibitem[{{Aoki} {et~al.}(2002){Aoki}, {Ando}, {Honda}, {Iye}, {Izumiura},
  {Kajino}, {Kambe}, {Kawanomonoto}, {Noguchi}, {Okita}, {Sadakane}, {Sato},
  {Shelton}, {Takada-Hidai}, {Takeda}, {Watanabe}, \& {Yoshida}}]{AOK02b}
{Aoki}, W., {Ando}, H., {Honda}, S., {et~al.} 2002, \pasj, 54, 427

\bibitem[{{Aoki} {et~al.}(2006){Aoki}, {Frebel}, {Christlieb}, {Norris},
  {Beers}, {Minezaki}, {Barklem}, {Honda}, {Takada-Hidai}, {Asplund}, {Ryan},
  {Tsangarides}, {Eriksson}, {Steinhauer}, {Deliyannis}, {Nomoto}, {Fujimoto},
  {Ando}, {Yoshii}, \& {Kajino}}]{AOK06b}
{Aoki}, W., {Frebel}, A., {Christlieb}, N., {et~al.} 2006, \apj, 639, 897

\bibitem[{{Aoki} {et~al.}(2005){Aoki}, {Honda}, {Beers}, {Kajino}, {Ando},
  {Norris}, {Ryan}, {Izumiura}, {Sadakane}, \& {Takada-Hidai}}]{AOK05}
{Aoki}, W., {Honda}, S., {Beers}, T.~C., {et~al.} 2005, \apj, 632, 611

\bibitem[{{Aoki} {et~al.}(2007){Aoki}, {Honda}, {Beers}, {Takada-Hidai},
  {Iwamoto}, {Tominaga}, {Umeda}, {Nomoto}, {Norris}, \& {Ryan}}]{AOK07c}
{Aoki}, W., {Honda}, S., {Beers}, T.~C., {et~al.} 2007, \apj, 660, 747

\bibitem[{{Arcones} {et~al.}(2007){Arcones}, {Janka}, \& {Scheck}}]{AJS07}
{Arcones}, A., {Janka}, H.-T., \& {Scheck}, L. 2007, \aap, 467, 1227

\bibitem[{{Arcones} \& {Montes}(2011)}]{ArcoMonte11}
{Arcones}, A. \& {Montes}, F. 2011, \apj, 731, 5

\bibitem[{{Argast} {et~al.}(2002){Argast}, {Samland}, {Thielemann}, \&
  {Gerhard}}]{Argast02}
{Argast}, D., {Samland}, M., {Thielemann}, F.-K., \& {Gerhard}, O.~E. 2002,
  \aap, 388, 842

\bibitem[{{Argast} {et~al.}(2004){Argast}, {Samland}, {Thielemann}, \&
  {Qian}}]{Argast04}
{Argast}, D., {Samland}, M., {Thielemann}, F.-K., \& {Qian}, Y.-Z. 2004, \aap,
  416, 997

\bibitem[{{Arlandini} {et~al.}(1999){Arlandini}, {K{\"a}ppeler}, {Wisshak},
  {Gallino}, {Lugaro}, {Busso}, \& {Straniero}}]{Arla99}
{Arlandini}, C., {K{\"a}ppeler}, F., {Wisshak}, K., {et~al.} 1999, \apj, 525,
  886

\bibitem[{{Arnould} {et~al.}(2007){Arnould}, {Goriely}, \& {Takahashi}}]{AGT07}
{Arnould}, M., {Goriely}, S., \& {Takahashi}, K. 2007, \physrep, 450, 97

\bibitem[{{Balcells} {et~al.}(2010){Balcells}, {Benn}, {Carter}, {Dalton},
  {Trager}, {Feltzing}, {Verheijen}, {Jarvis}, {Percival}, {Abrams}, {Agocs},
  {Brown}, {Cano}, {Evans}, {Helmi}, {Lewis}, {McLure}, {Peletier},
  {P{\'e}rez-Fournon}, {Sharples}, {Tosh}, {Trujillo}, {Walton}, \&
  {Westhall}}]{weave}
{Balcells}, M., {Benn}, C.~R., {Carter}, D., {et~al.} 2010, in Society of
  Photo-Optical Instrumentation Engineers (SPIE) Conference Series, Vol. 7735,
  Society of Photo-Optical Instrumentation Engineers (SPIE) Conference Series

\bibitem[{{Barbuy} {et~al.}(2009){Barbuy}, {Zoccali}, {Ortolani}, {Hill},
  {Minniti}, {Bica}, {Renzini}, \& {G{\'o}mez}}]{Barbuy09}
{Barbuy}, B., {Zoccali}, M., {Ortolani}, S., {et~al.} 2009, \aap, 507, 405

\bibitem[{{Barklem} {et~al.}(2005){Barklem}, {Christlieb}, {Beers}, {Hill},
  {Bessell}, {Holmberg}, {Marsteller}, {Rossi}, {Zickgraf}, \&
  {Reimers}}]{BAR05}
{Barklem}, P.~S., {Christlieb}, N., {Beers}, T.~C., {et~al.} 2005, \aap, 439,
  129

\bibitem[{{Bisterzo} {et~al.}(2012){Bisterzo}, {Gallino}, {Straniero},
  {Cristallo}, \& {K{\"a}ppeler}}]{Bisterzo12}
{Bisterzo}, S., {Gallino}, R., {Straniero}, O., {Cristallo}, S., \&
  {K{\"a}ppeler}, F. 2012, \mnras, 422, 849

\bibitem[{{Bonifacio} {et~al.}(2012){Bonifacio}, {Sbordone}, {Caffau},
  {Ludwig}, {Spite}, {Gonz{\'a}lez Hern{\'a}ndez}, \& {Behara}}]{Boni12}
{Bonifacio}, P., {Sbordone}, L., {Caffau}, E., {et~al.} 2012, \aap, 542, A87

\bibitem[{{Bonifacio} {et~al.}(2009){Bonifacio}, {Spite}, {Cayrel}, {Hill},
  {Spite}, {Fran{\c c}ois}, {Plez}, {Ludwig}, {Caffau}, {Molaro}, {Depagne},
  {Andersen}, {Barbuy}, {Beers}, {Nordstr{\"o}m}, \& {Primas}}]{BON09}
{Bonifacio}, P., {Spite}, M., {Cayrel}, R., {et~al.} 2009, \aap, 501, 519

\bibitem[{{Boyd} {et~al.}(2012){Boyd}, {Famiano}, {Meyer}, {Motizuki},
  {Kajino}, \& {Roederer}}]{Boyd12}
{Boyd}, R.~N., {Famiano}, M.~A., {Meyer}, B.~S., {et~al.} 2012, \apjl, 744, L14

\bibitem[{{Bromm} {et~al.}(1999){Bromm}, {Coppi}, \& {Larson}}]{Bromm99}
{Bromm}, V., {Coppi}, P.~S., \& {Larson}, R.~B. 1999, \apjl, 527, L5

\bibitem[{{Bromm} \& {Larson}(2004)}]{BrommL04}
{Bromm}, V. \& {Larson}, R.~B. 2004, \araa, 42, 79

\bibitem[{{Bromm} \& {Yoshida}(2011)}]{BrommY11}
{Bromm}, V. \& {Yoshida}, N. 2011, \araa, 49, 373

\bibitem[{Burbidge {et~al.}(1957)Burbidge, Burbidge, Fowler, \& Hoyle}]{BBFH57}
Burbidge, E.~M., Burbidge, G.~R., Fowler, W.~A., \& Hoyle, F. 1957, Rev. Mod.
  Phys., 29, 547

\bibitem[{{Caffau} {et~al.}(2011){Caffau}, {Bonifacio}, {Fran{\c c}ois},
  {Sbordone}, {Monaco}, {Spite}, {Spite}, {Ludwig}, {Cayrel}, {Zaggia},
  {Hammer}, {Randich}, {Molaro}, \& {Hill}}]{Caffau11}
{Caffau}, E., {Bonifacio}, P., {Fran{\c c}ois}, P., {et~al.} 2011, \nat, 477,
  67

\bibitem[{{Cameron}(2003)}]{Cam2003}
{Cameron}, A.~G.~W. 2003, \apj, 587, 327

\bibitem[{{Cantiello} {et~al.}(2007){Cantiello}, {Yoon}, {Langer}, \&
  {Livio}}]{Cantiello07}
{Cantiello}, M., {Yoon}, S.-C., {Langer}, N., \& {Livio}, M. 2007, \aap, 465,
  L29

\bibitem[{{Caughlan} \& {Fowler}(1988)}]{CF88}
{Caughlan}, G.~R. \& {Fowler}, W.~A. 1988, Atomic Data and Nuclear Data Tables,
  40, 283

\bibitem[{{Cayrel} {et~al.}(2004){Cayrel}, {Depagne}, {Spite}, {Hill}, {Spite},
  {Fran{\c c}ois}, {Plez}, {Beers}, {Primas}, {Andersen}, {Barbuy},
  {Bonifacio}, {Molaro}, \& {Nordstr{\"o}m}}]{Cayrel04}
{Cayrel}, R., {Depagne}, E., {Spite}, M., {et~al.} 2004, \aap, 416, 1117

\bibitem[{{Cayrel} {et~al.}(2001){Cayrel}, {Hill}, {Beers}, {Barbuy}, {Spite},
  {Spite}, {Plez}, {Andersen}, {Bonifacio}, {Fran{\c c}ois}, {Molaro},
  {Nordstr{\"o}m}, \& {Primas}}]{Cayrel01}
{Cayrel}, R., {Hill}, V., {Beers}, T.~C., {et~al.} 2001, \nat, 409, 691

\bibitem[{{Cescutti}(2008)}]{Cesc08}
{Cescutti}, G. 2008, \aap, 481, 691

\bibitem[{{Cescutti} \& {Chiappini}(2010)}]{Cesc10}
{Cescutti}, G. \& {Chiappini}, C. 2010, \aap, 515, A102

\bibitem[{{Cescutti} {et~al.}(2006){Cescutti}, {Fran{\c c}ois}, {Matteucci},
  {Cayrel}, \& {Spite}}]{Cesc06}
{Cescutti}, G., {Fran{\c c}ois}, P., {Matteucci}, F., {Cayrel}, R., \& {Spite},
  M. 2006, \aap, 448, 557

\bibitem[{{Chiappini}(2012)}]{Chiappini11as}
{Chiappini}, C. 2012, {Red Giant Stars: Probing the Milky Way Chemical
  Enrichment}, ed. A.~{Miglio}, J.~{Montalb{\'a}n}, \& A.~{Noels}, 147

\bibitem[{{Chiappini} {et~al.}(2008){Chiappini}, {Ekstr{\"o}m}, {Meynet},
  {Hirschi}, {Maeder}, \& {Charbonnel}}]{Chiappini08}
{Chiappini}, C., {Ekstr{\"o}m}, S., {Meynet}, G., {et~al.} 2008, \aap, 479, L9

\bibitem[{{Chiappini} {et~al.}(2011){Chiappini}, {Frischknecht}, {Meynet},
  {Hirschi}, {Barbuy}, {Pignatari}, {Decressin}, \& {Maeder}}]{Chiappini11}
{Chiappini}, C., {Frischknecht}, U., {Meynet}, G., {et~al.} 2011, \nat, 472,
  454

\bibitem[{{Chiappini} {et~al.}(2006){Chiappini}, {Hirschi}, {Meynet},
  {Ekstr{\"o}m}, {Maeder}, \& {Matteucci}}]{Chiappini06}
{Chiappini}, C., {Hirschi}, R., {Meynet}, G., {et~al.} 2006, \aap, 449, L27

\bibitem[{{Christlieb} {et~al.}(2002){Christlieb}, {Bessell}, {Beers},
  {Gustafsson}, {Korn}, {Barklem}, {Karlsson}, {Mizuno-Wiedner}, \&
  {Rossi}}]{Christlieb02}
{Christlieb}, N., {Bessell}, M.~S., {Beers}, T.~C., {et~al.} 2002, \nat, 419,
  904

\bibitem[{{Christlieb} {et~al.}(2008){Christlieb}, {Sch{\"o}rck}, {Frebel},
  {Beers}, {Wisotzki}, \& {Reimers}}]{Christlieb08}
{Christlieb}, N., {Sch{\"o}rck}, T., {Frebel}, A., {et~al.} 2008, \aap, 484,
  721

\bibitem[{{Clark} {et~al.}(2011){Clark}, {Glover}, {Smith}, {Greif}, {Klessen},
  \& {Bromm}}]{Clark11}
{Clark}, P.~C., {Glover}, S.~C.~O., {Smith}, R.~J., {et~al.} 2011, Science,
  331, 1040

\bibitem[{{Cohen} {et~al.}(2008){Cohen}, {Christlieb}, {McWilliam}, {Shectman},
  {Thompson}, {Melendez}, {Wisotzki}, \& {Reimers}}]{COH08}
{Cohen}, J.~G., {Christlieb}, N., {McWilliam}, A., {et~al.} 2008, \apj, 672,
  320

\bibitem[{{Cowan} {et~al.}(2002){Cowan}, {Sneden}, {Burles}, {Ivans}, {Beers},
  {Truran}, {Lawler}, {Primas}, {Fuller}, {Pfeiffer}, \& {Kratz}}]{COW02}
{Cowan}, J.~J., {Sneden}, C., {Burles}, S., {et~al.} 2002, \apj, 572, 861

\bibitem[{{de Jong} {et~al.}(2012){de Jong}, {Bellido-Tirado}, {Chiappini},
  {Depagne}, {Haynes}, {Johl}, {Schnurr}, {Schwope}, {Walcher}, {Dionies},
  {Haynes}, {Kelz}, {Kitaura}, {Lamer}, {Minchev}, {M{\"u}ller}, {Nuza},
  {Olaya}, {Piffl}, {Popow}, {Steinmetz}, {Ural}, {Williams}, {Winkler},
  {Wisotzki}, {Ansorge}, {Banerji}, {Gonzalez Solares}, {Irwin}, {Kennicutt},
  {King}, {McMahon}, {Koposov}, {Parry}, {Sun}, {Walton}, {Finger}, {Iwert},
  {Krumpe}, {Lizon}, {Vincenzo}, {Amans}, {Bonifacio}, {Cohen}, {Francois},
  {Jagourel}, {Mignot}, {Royer}, {Sartoretti}, {Bender}, {Grupp}, {Hess},
  {Lang-Bardl}, {Muschielok}, {B{\"o}hringer}, {Boller}, {Bongiorno}, {Brusa},
  {Dwelly}, {Merloni}, {Nandra}, {Salvato}, {Pragt}, {Navarro}, {Gerlofsma},
  {Roelfsema}, {Dalton}, {Middleton}, {Tosh}, {Boeche}, {Caffau}, {Christlieb},
  {Grebel}, {Hansen}, {Koch}, {Ludwig}, {Quirrenbach}, {Sbordone}, {Seifert},
  {Thimm}, {Trifonov}, {Helmi}, {Trager}, {Feltzing}, {Korn}, \&
  {Boland}}]{4most}
{de Jong}, R.~S., {Bellido-Tirado}, O., {Chiappini}, C., {et~al.} 2012, in
  Society of Photo-Optical Instrumentation Engineers (SPIE) Conference Series,
  Vol. 8446, Society of Photo-Optical Instrumentation Engineers (SPIE)
  Conference Series

\bibitem[{{Eisenstein} {et~al.}(2011){Eisenstein}, {Weinberg}, {Agol},
  {Aihara}, {Allende Prieto}, {Anderson}, {Arns}, {Aubourg}, {Bailey},
  {Balbinot}, \& et~al.}]{apogee}
{Eisenstein}, D.~J., {Weinberg}, D.~H., {Agol}, E., {et~al.} 2011, \aj, 142, 72

\bibitem[{{Fabbian} {et~al.}(2009){Fabbian}, {Nissen}, {Asplund}, {Pettini}, \&
  {Akerman}}]{Fabbian}
{Fabbian}, D., {Nissen}, P.~E., {Asplund}, M., {Pettini}, M., \& {Akerman}, C.
  2009, \aap, 500, 1143

\bibitem[{{Farouqi} {et~al.}(2009){Farouqi}, {Kratz}, {Mashonkina}, {Pfeiffer},
  {Cowan}, {Thielemann}, \& {Truran}}]{Farouqi09}
{Farouqi}, K., {Kratz}, K.-L., {Mashonkina}, L.~I., {et~al.} 2009, \apjl, 694,
  L49

\bibitem[{{Ferrara}(2012)}]{Ferrara12}
{Ferrara}, A. 2012, in American Institute of Physics Conference Series, Vol.
  1480, American Institute of Physics Conference Series, ed. M.~{Umemura} \&
  K.~{Omukai}, 317--324

\bibitem[{{Fran{\c c}ois} {et~al.}(2007){Fran{\c c}ois}, {Depagne}, {Hill},
  {Spite}, {Spite}, {Plez}, {Beers}, {Andersen}, {James}, {Barbuy}, {Cayrel},
  {Bonifacio}, {Molaro}, {Nordstr{\"o}m}, \& {Primas}}]{Franc07}
{Fran{\c c}ois}, P., {Depagne}, E., {Hill}, V., {et~al.} 2007, \aap, 476, 935

\bibitem[{{Fran{\c c}ois} {et~al.}(2004){Fran{\c c}ois}, {Matteucci}, {Cayrel},
  {Spite}, {Spite}, \& {Chiappini}}]{Franc04}
{Fran{\c c}ois}, P., {Matteucci}, F., {Cayrel}, R., {et~al.} 2004, \aap, 421,
  613

\bibitem[{{Frebel}(2010)}]{Frebel10}
{Frebel}, A. 2010, Astronomische Nachrichten, 331, 474

\bibitem[{{Frebel} {et~al.}(2005){Frebel}, {Aoki}, {Christlieb}, {Ando},
  {Asplund}, {Barklem}, {Beers}, {Eriksson}, {Fechner}, {Fujimoto}, {Honda},
  {Kajino}, {Minezaki}, {Nomoto}, {Norris}, {Ryan}, {Takada-Hidai},
  {Tsangarides}, \& {Yoshii}}]{Frebel05}
{Frebel}, A., {Aoki}, W., {Christlieb}, N., {et~al.} 2005, \nat, 434, 871

\bibitem[{{Frebel} {et~al.}(2008){Frebel}, {Collet}, {Eriksson}, {Christlieb},
  \& {Aoki}}]{Frebel08}
{Frebel}, A., {Collet}, R., {Eriksson}, K., {Christlieb}, N., \& {Aoki}, W.
  2008, \apj, 684, 588

\bibitem[{{Freeman} \& {Bland-Hawthorn}(2002)}]{Freeman}
{Freeman}, K. \& {Bland-Hawthorn}, J. 2002, \araa, 40, 487

\bibitem[{{Freiburghaus} {et~al.}(1999){Freiburghaus}, {Rosswog}, \&
  {Thielemann}}]{FRT99}
{Freiburghaus}, C., {Rosswog}, S., \& {Thielemann}, F.-K. 1999, \apjl, 525,
  L121

\bibitem[{{Frischknecht} {et~al.}(2012){Frischknecht}, {Hirschi}, \&
  {Thielemann}}]{Frisch12}
{Frischknecht}, U., {Hirschi}, R., \& {Thielemann}, F.-K. 2012, \aap, 538, L2

\bibitem[{{Gallagher} {et~al.}(2010){Gallagher}, {Ryan}, {Garc{\'{\i}}a
  P{\'e}rez}, \& {Aoki}}]{Gallagher10}
{Gallagher}, A.~J., {Ryan}, S.~G., {Garc{\'{\i}}a P{\'e}rez}, A.~E., \& {Aoki},
  W. 2010, \aap, 523, A24

\bibitem[{{Gallagher} {et~al.}(2012){Gallagher}, {Ryan}, {Hosford},
  {Garc{\'{\i}}a P{\'e}rez}, {Aoki}, \& {Honda}}]{Gallagher12}
{Gallagher}, A.~J., {Ryan}, S.~G., {Hosford}, A., {et~al.} 2012, \aap, 538,
  A118

\bibitem[{{Goriely} {et~al.}(2011){Goriely}, {Bauswein}, \&
  {Janka}}]{Goriely11}
{Goriely}, S., {Bauswein}, A., \& {Janka}, H.-T. 2011, \apjl, 738, L32

\bibitem[{{Greif} {et~al.}(2012){Greif}, {Bromm}, {Clark}, {Glover}, {Smith},
  {Klessen}, {Yoshida}, \& {Springel}}]{Greif12}
{Greif}, T.~H., {Bromm}, V., {Clark}, P.~C., {et~al.} 2012, \mnras, 424, 399

\bibitem[{{Greif} {et~al.}(2011){Greif}, {Springel}, {White}, {Glover},
  {Clark}, {Smith}, {Klessen}, \& {Bromm}}]{Greif11}
{Greif}, T.~H., {Springel}, V., {White}, S.~D.~M., {et~al.} 2011, \apj, 737, 75

\bibitem[{{Grevesse} {et~al.}(2010){Grevesse}, {Asplund}, {Sauval}, \&
  {Scott}}]{Grevesse10}
{Grevesse}, N., {Asplund}, M., {Sauval}, A.~J., \& {Scott}, P. 2010, \apss,
  328, 179

\bibitem[{{Hansen} {et~al.}(2013){Hansen}, {Bergemann}, {Cescutti}, {Fran{\c
  c}ois}, {Arcones}, {Karakas}, {Lind}, \& {Chiappini}}]{Hansen13}
{Hansen}, C.~J., {Bergemann}, M., {Cescutti}, G., {et~al.} 2013, \aap, 551, A57

\bibitem[{{Hayek} {et~al.}(2009){Hayek}, {Wiesendahl}, {Christlieb},
  {Eriksson}, {Korn}, {Barklem}, {Hill}, {Beers}, {Farouqi}, {Pfeiffer}, \&
  {Kratz}}]{HAY09}
{Hayek}, W., {Wiesendahl}, U., {Christlieb}, N., {et~al.} 2009, \aap, 504, 511

\bibitem[{{Honda} {et~al.}(2004){Honda}, {Aoki}, {Kajino}, {Ando}, {Beers},
  {Izumiura}, {Sadakane}, \& {Takada-Hidai}}]{Honda04}
{Honda}, S., {Aoki}, W., {Kajino}, T., {et~al.} 2004, \apj, 607, 474

\bibitem[{{Ishimaru} \& {Wanajo}(1999)}]{Ishimaru99}
{Ishimaru}, Y. \& {Wanajo}, S. 1999, \apjl, 511, L33

\bibitem[{{Ivans} {et~al.}(2006){Ivans}, {Simmerer}, {Sneden}, {Lawler},
  {Cowan}, {Gallino}, \& {Bisterzo}}]{IVA06}
{Ivans}, I.~I., {Simmerer}, J., {Sneden}, C., {et~al.} 2006, \apj, 645, 613

\bibitem[{{Ivans} {et~al.}(2003){Ivans}, {Sneden}, {James}, {Preston},
  {Fulbright}, {H{\"o}flich}, {Carney}, \& {Wheeler}}]{IVA03}
{Ivans}, I.~I., {Sneden}, C., {James}, C.~R., {et~al.} 2003, \apj, 592, 906

\bibitem[{{Jaikumar} {et~al.}(2007){Jaikumar}, {Meyer}, {Otsuki}, \&
  {Ouyed}}]{JMO07}
{Jaikumar}, P., {Meyer}, B.~S., {Otsuki}, K., \& {Ouyed}, R. 2007, \aap, 471,
  227

\bibitem[{{Joggerst} \& {Whalen}(2011)}]{JoggW11}
{Joggerst}, C.~C. \& {Whalen}, D.~J. 2011, \apj, 728, 129

\bibitem[{{K{\"a}ppeler} {et~al.}(2011){K{\"a}ppeler}, {Gallino}, {Bisterzo},
  \& {Aoki}}]{Kaeppeler}
{K{\"a}ppeler}, F., {Gallino}, R., {Bisterzo}, S., \& {Aoki}, W. 2011, Reviews
  of Modern Physics, 83, 157

\bibitem[{{Karlsson} {et~al.}(2011){Karlsson}, {Bromm}, \&
  {Bland-Hawthorn}}]{Karlsson11}
{Karlsson}, T., {Bromm}, V., \& {Bland-Hawthorn}, J. 2011, ArXiv e-prints

\bibitem[{{Karlsson} \& {Gustafsson}(2005)}]{KaGu05}
{Karlsson}, T. \& {Gustafsson}, B. 2005, \aap, 436, 879

\bibitem[{{Keller} {et~al.}(2007){Keller}, {Schmidt}, {Bessell}, {Conroy},
  {Francis}, {Granlund}, {Kowald}, {Oates}, {Martin-Jones}, {Preston},
  {Tisserand}, {Vaccarella}, \& {Waterson}}]{skymapper}
{Keller}, S.~C., {Schmidt}, B.~P., {Bessell}, M.~S., {et~al.} 2007, \pasa, 24,
  1

\bibitem[{{Kitaura} {et~al.}(2006){Kitaura}, {Janka}, \& {Hillebrandt}}]{KJH06}
{Kitaura}, F.~S., {Janka}, H.-T., \& {Hillebrandt}, W. 2006, \aap, 450, 345

\bibitem[{{Kobayashi} {et~al.}(2011){Kobayashi}, {Karakas}, \&
  {Umeda}}]{Kobayashi11}
{Kobayashi}, C., {Karakas}, A.~I., \& {Umeda}, H. 2011, \mnras, 414, 3231

\bibitem[{{Korobkin} {et~al.}(2012){Korobkin}, {Rosswog}, {Arcones}, \&
  {Winteler}}]{Korobkin12}
{Korobkin}, O., {Rosswog}, S., {Arcones}, A., \& {Winteler}, C. 2012, \mnras,
  426, 1940

\bibitem[{{Lai} {et~al.}(2008){Lai}, {Bolte}, {Johnson}, {Lucatello}, {Heger},
  \& {Woosley}}]{LAI08}
{Lai}, D.~K., {Bolte}, M., {Johnson}, J.~A., {et~al.} 2008, \apj, 681, 1524

\bibitem[{{Lai} {et~al.}(2007){Lai}, {Johnson}, {Bolte}, \&
  {Lucatello}}]{LAI07}
{Lai}, D.~K., {Johnson}, J.~A., {Bolte}, M., \& {Lucatello}, S. 2007, \apj,
  667, 1185

\bibitem[{{Li} {et~al.}(2010){Li}, {Christlieb}, {Sch{\"o}rck}, {Norris},
  {Bessell}, {Yong}, {Beers}, {Lee}, {Frebel}, \& {Zhao}}]{Li10}
{Li}, H.~N., {Christlieb}, N., {Sch{\"o}rck}, T., {et~al.} 2010, \aap, 521, A10

\bibitem[{{Lucatello} {et~al.}(2006){Lucatello}, {Beers}, {Christlieb},
  {Barklem}, {Rossi}, {Marsteller}, {Sivarani}, \& {Lee}}]{Lucatello06}
{Lucatello}, S., {Beers}, T.~C., {Christlieb}, N., {et~al.} 2006, \apjl, 652,
  L37

\bibitem[{{Lugaro} {et~al.}(2012){Lugaro}, {Karakas}, {Stancliffe}, \&
  {Rijs}}]{Lugaro12}
{Lugaro}, M., {Karakas}, A.~I., {Stancliffe}, R.~J., \& {Rijs}, C. 2012, \apj,
  747, 2

\bibitem[{{Maeder} \& {Meynet}(1989)}]{MM89}
{Maeder}, A. \& {Meynet}, G. 1989, \aap, 210, 155

\bibitem[{{Maeder} \& {Meynet}(2012)}]{MM12}
{Maeder}, A. \& {Meynet}, G. 2012, Reviews of Modern Physics, 84, 25

\bibitem[{{Masseron} {et~al.}(2010){Masseron}, {Johnson}, {Plez}, {van Eck},
  {Primas}, {Goriely}, \& {Jorissen}}]{Masseron10}
{Masseron}, T., {Johnson}, J.~A., {Plez}, B., {et~al.} 2010, \aap, 509, A93

\bibitem[{{Masseron} {et~al.}(2006){Masseron}, {van Eck}, {Famaey}, {Goriely},
  {Plez}, {Siess}, {Beers}, {Primas}, \& {Jorissen}}]{MAS06}
{Masseron}, T., {van Eck}, S., {Famaey}, B., {et~al.} 2006, \aap, 455, 1059

\bibitem[{{McWilliam}(1998)}]{MCW98}
{McWilliam}, A. 1998, \aj, 115, 1640

\bibitem[{{McWilliam} {et~al.}(1995){McWilliam}, {Preston}, {Sneden}, \&
  {Searle}}]{MCW95}
{McWilliam}, A., {Preston}, G.~W., {Sneden}, C., \& {Searle}, L. 1995, \aj,
  109, 2757

\bibitem[{{Meynet} {et~al.}(2008){Meynet}, {Ekstr{\"o}m}, {Maeder}, {Hirschi},
  {Chiappini}, \& {Georgy}}]{Meynet08}
{Meynet}, G., {Ekstr{\"o}m}, S., {Maeder}, A., {et~al.} 2008, in American
  Institute of Physics Conference Series, Vol. 990, First Stars III, ed. B.~W.
  {O'Shea} \& A.~{Heger}, 212--216

\bibitem[{{Meynet} {et~al.}(2010){Meynet}, {Hirschi}, {Ekstrom}, {Maeder},
  {Georgy}, {Eggenberger}, \& {Chiappini}}]{Meynet10}
{Meynet}, G., {Hirschi}, R., {Ekstrom}, S., {et~al.} 2010, \aap, 521, A30

\bibitem[{{Montes} {et~al.}(2007){Montes}, {Beers}, {Cowan}, {Elliot},
  {Farouqi}, {Gallino}, {Heil}, {Kratz}, {Pfeiffer}, {Pignatari}, \&
  {Schatz}}]{Montes07}
{Montes}, F., {Beers}, T.~C., {Cowan}, J., {et~al.} 2007, \apj, 671, 1685

\bibitem[{{Muijres} {et~al.}(2012){Muijres}, {Vink}, {de Koter}, {Hirschi},
  {Langer}, \& {Yoon}}]{Muy}
{Muijres}, L., {Vink}, J.~S., {de Koter}, A., {et~al.} 2012, \aap, 546, A42

\bibitem[{{Norris} {et~al.}(2013){Norris}, {Yong}, {Bessell}, {Christlieb},
  {Asplund}, {Gilmore}, {Wyse}, {Beers}, {Barklem}, {Frebel}, \&
  {Ryan}}]{Norris13}
{Norris}, J.~E., {Yong}, D., {Bessell}, M.~S., {et~al.} 2013, \apj, 762, 28

\bibitem[{{Pignatari} {et~al.}(2008){Pignatari}, {Gallino}, {Meynet},
  {Hirschi}, {Herwig}, \& {Wiescher}}]{Pigna08}
{Pignatari}, M., {Gallino}, R., {Meynet}, G., {et~al.} 2008, \apjl, 687, L95

\bibitem[{{Prantzos}(2012)}]{Prantzos12}
{Prantzos}, N. 2012, \aap, 542, A67

\bibitem[{{Preston} {et~al.}(2006){Preston}, {Sneden}, {Thompson}, {Shectman},
  \& {Burley}}]{PRE06}
{Preston}, G.~W., {Sneden}, C., {Thompson}, I.~B., {Shectman}, S.~A., \&
  {Burley}, G.~S. 2006, \aj, 132, 85

\bibitem[{{Raiteri} {et~al.}(1992){Raiteri}, {Gallino}, \& {Busso}}]{Raiteri92}
{Raiteri}, C.~M., {Gallino}, R., \& {Busso}, M. 1992, \apj, 387, 263

\bibitem[{{Roederer}(2013)}]{Roederer13}
{Roederer}, I.~U. 2013, \aj, 145, 26

\bibitem[{{Roederer} {et~al.}(2010){Roederer}, {Cowan}, {Karakas}, {Kratz},
  {Lugaro}, {Simmerer}, {Farouqi}, \& {Sneden}}]{Roederer10}
{Roederer}, I.~U., {Cowan}, J.~J., {Karakas}, A.~I., {et~al.} 2010, \apj, 724,
  975

\bibitem[{{Roederer} {et~al.}(2008){Roederer}, {Frebel}, {Shetrone}, {Allende
  Prieto}, {Rhee}, {Gallino}, {Bisterzo}, {Sneden}, {Beers}, \&
  {Cowan}}]{ROE08}
{Roederer}, I.~U., {Frebel}, A., {Shetrone}, M.~D., {et~al.} 2008, \apj, 679,
  1549

\bibitem[{{Scalo}(1986)}]{Scalo86}
{Scalo}, J.~M. 1986, \fcp, 11, 1

\bibitem[{{Sch{\"o}rck} {et~al.}(2009){Sch{\"o}rck}, {Christlieb}, {Cohen},
  {Beers}, {Shectman}, {Thompson}, {McWilliam}, {Bessell}, {Norris},
  {Mel{\'e}ndez}, {Ram{\'{\i}}rez}, {Haynes}, {Cass}, {Hartley}, {Russell},
  {Watson}, {Zickgraf}, {Behnke}, {Fechner}, {Fuhrmeister}, {Barklem},
  {Edvardsson}, {Frebel}, {Wisotzki}, \& {Reimers}}]{Schoerck09}
{Sch{\"o}rck}, T., {Christlieb}, N., {Cohen}, J.~G., {et~al.} 2009, \aap, 507,
  817

\bibitem[{{Siess}(2007)}]{Siess07}
{Siess}, L. 2007, \aap, 476, 893

\bibitem[{{Siess}(2010)}]{Siess10}
{Siess}, L. 2010, \aap, 512, A10

\bibitem[{{Simmerer} {et~al.}(2004){Simmerer}, {Sneden}, {Cowan}, {Collier},
  {Woolf}, \& {Lawler}}]{SSC04}
{Simmerer}, J., {Sneden}, C., {Cowan}, J.~J., {et~al.} 2004, \apj, 617, 1091

\bibitem[{{Sneden} {et~al.}(2008){Sneden}, {Cowan}, \& {Gallino}}]{Sneden08}
{Sneden}, C., {Cowan}, J.~J., \& {Gallino}, R. 2008, \araa, 46, 241

\bibitem[{{Sneden} {et~al.}(2003){Sneden}, {Cowan}, {Lawler}, {Ivans},
  {Burles}, {Beers}, {Primas}, {Hill}, {Truran}, {Fuller}, {Pfeiffer}, \&
  {Kratz}}]{Sneden03}
{Sneden}, C., {Cowan}, J.~J., {Lawler}, J.~E., {et~al.} 2003, \apj, 591, 936

\bibitem[{{Sneden} {et~al.}(1996){Sneden}, {McWilliam}, {Preston}, {Cowan},
  {Burris}, \& {Armosky}}]{Sneden96}
{Sneden}, C., {McWilliam}, A., {Preston}, G.~W., {et~al.} 1996, \apj, 467, 819

\bibitem[{{Spite} {et~al.}(2006){Spite}, {Cayrel}, {Hill}, {Spite}, {Fran{\c
  c}ois}, {Plez}, {Bonifacio}, {Molaro}, {Depagne}, {Andersen}, {Barbuy},
  {Beers}, {Nordstr{\"o}m}, \& {Primas}}]{Spite06}
{Spite}, M., {Cayrel}, R., {Hill}, V., {et~al.} 2006, \aap, 455, 291

\bibitem[{{Spite} {et~al.}(2005){Spite}, {Cayrel}, {Plez}, {Hill}, {Spite},
  {Depagne}, {Fran{\c c}ois}, {Bonifacio}, {Barbuy}, {Beers}, {Andersen},
  {Molaro}, {Nordstr{\"o}m}, \& {Primas}}]{Spite05}
{Spite}, M., {Cayrel}, R., {Plez}, B., {et~al.} 2005, \aap, 430, 655

\bibitem[{{Stacy} {et~al.}(2011){Stacy}, {Bromm}, \& {Loeb}}]{Stacy11}
{Stacy}, A., {Bromm}, V., \& {Loeb}, A. 2011, \mnras, 413, 543

\bibitem[{{Stacy} {et~al.}(2012){Stacy}, {Greif}, {Klessen}, {Bromm}, \&
  {Loeb}}]{Stacy12}
{Stacy}, A., {Greif}, T.~H., {Klessen}, R.~S., {Bromm}, V., \& {Loeb}, A. 2012,
  ArXiv e-prints

\bibitem[{{Takahashi} {et~al.}(1994){Takahashi}, {Witti}, \& {Janka}}]{TWJ94}
{Takahashi}, K., {Witti}, J., \& {Janka}, H.-T. 1994, \aap, 286, 857

\bibitem[{{Thielemann} {et~al.}(2011){Thielemann}, {Arcones}, {K{\"a}ppeli},
  {Liebend{\"o}rfer}, {Rauscher}, {Winteler}, {Fr{\"o}hlich}, {Dillmann},
  {Fischer}, {Martinez-Pinedo}, {Langanke}, {Farouqi}, {Kratz}, {Panov}, \&
  {Korneev}}]{Thielemann11}
{Thielemann}, F.-K., {Arcones}, A., {K{\"a}ppeli}, R., {et~al.} 2011, Progress
  in Particle and Nuclear Physics, 66, 346

\bibitem[{{Thompson}(2003)}]{Thom03}
{Thompson}, T.~A. 2003, \apjl, 585, L33

\bibitem[{{Thornton} {et~al.}(1998){Thornton}, {Gaudlitz}, {Janka}, \&
  {Steinmetz}}]{Thornton98}
{Thornton}, K., {Gaudlitz}, M., {Janka}, H.-T., \& {Steinmetz}, M. 1998, \apj,
  500, 95

\bibitem[{{Travaglio} {et~al.}(2001){Travaglio}, {Galli}, \&
  {Burkert}}]{Trava01}
{Travaglio}, C., {Galli}, D., \& {Burkert}, A. 2001, \apj, 547, 217

\bibitem[{{Travaglio} {et~al.}(1999){Travaglio}, {Galli}, {Gallino}, {Busso},
  {Ferrini}, \& {Straniero}}]{Trava99}
{Travaglio}, C., {Galli}, D., {Gallino}, R., {et~al.} 1999, \apj, 521, 691

\bibitem[{{Travaglio} {et~al.}(2004){Travaglio}, {Gallino}, {Arnone}, {Cowan},
  {Jordan}, \& {Sneden}}]{Trava04}
{Travaglio}, C., {Gallino}, R., {Arnone}, E., {et~al.} 2004, \apj, 601, 864

\bibitem[{{Truran}(1981)}]{Truran81}
{Truran}, J.~W. 1981, \aap, 97, 391

\bibitem[{{Tsujimoto} {et~al.}(1999){Tsujimoto}, {Shigeyama}, \&
  {Yoshii}}]{Tsujimoto99}
{Tsujimoto}, T., {Shigeyama}, T., \& {Yoshii}, Y. 1999, \apjl, 519, L63

\bibitem[{{Tumlinson}(2010)}]{Tum10}
{Tumlinson}, J. 2010, \apj, 708, 1398

\bibitem[{{Tur} {et~al.}(2009){Tur}, {Heger}, \& {Austin}}]{Tur09}
{Tur}, C., {Heger}, A., \& {Austin}, S.~M. 2009, \apj, 702, 1068

\bibitem[{{Wanajo} {et~al.}(2011){Wanajo}, {Janka}, \& {M{\"u}ller}}]{WJM11}
{Wanajo}, S., {Janka}, H.-T., \& {M{\"u}ller}, B. 2011, \apjl, 726, L15

\bibitem[{{Wanajo} {et~al.}(2009){Wanajo}, {Nomoto}, {Janka}, {Kitaura}, \&
  {M{\"u}ller}}]{WNJ09}
{Wanajo}, S., {Nomoto}, K., {Janka}, H.-T., {Kitaura}, F.~S., \& {M{\"u}ller},
  B. 2009, \apj, 695, 208

\bibitem[{{Wasserburg} {et~al.}(1996){Wasserburg}, {Busso}, \&
  {Gallino}}]{Wasserburg96}
{Wasserburg}, G.~J., {Busso}, M., \& {Gallino}, R. 1996, \apjl, 466, L109

\bibitem[{{Westin} {et~al.}(2000){Westin}, {Sneden}, {Gustafsson}, \&
  {Cowan}}]{WES00}
{Westin}, J., {Sneden}, C., {Gustafsson}, B., \& {Cowan}, J.~J. 2000, \apj,
  530, 783

\bibitem[{{Winteler} {et~al.}(2012){Winteler}, {K{\"a}ppeli}, {Perego},
  {Arcones}, {Vasset}, {Nishimura}, {Liebend{\"o}rfer}, \&
  {Thielemann}}]{Winteler12}
{Winteler}, C., {K{\"a}ppeli}, R., {Perego}, A., {et~al.} 2012, \apjl, 750, L22

\bibitem[{{Woosley} {et~al.}(2002){Woosley}, {Heger}, \& {Weaver}}]{Woosley02}
{Woosley}, S.~E., {Heger}, A., \& {Weaver}, T.~A. 2002, Reviews of Modern
  Physics, 74, 1015

\bibitem[{{Woosley} {et~al.}(1994){Woosley}, {Wilson}, {Mathews}, {Hoffman}, \&
  {Meyer}}]{WWM94}
{Woosley}, S.~E., {Wilson}, J.~R., {Mathews}, G.~J., {Hoffman}, R.~D., \&
  {Meyer}, B.~S. 1994, \apj, 433, 229

\bibitem[{{Yanny} {et~al.}(2009){Yanny}, {Rockosi}, {Newberg}, {Knapp},
  {Adelman-McCarthy}, {Alcorn}, {Allam}, {Allende Prieto}, {An}, {Anderson},
  {Anderson}, {Bailer-Jones}, {Bastian}, {Beers}, {Bell}, {Belokurov},
  {Bizyaev}, {Blythe}, {Bochanski}, {Boroski}, {Brinchmann}, {Brinkmann},
  {Brewington}, {Carey}, {Cudworth}, {Evans}, {Evans}, {Gates}, {G{\"a}nsicke},
  {Gillespie}, {Gilmore}, {Nebot Gomez-Moran}, {Grebel}, {Greenwell}, {Gunn},
  {Jordan}, {Jordan}, {Harding}, {Harris}, {Hendry}, {Holder}, {Ivans},
  {Ivezi{\v c}}, {Jester}, {Johnson}, {Kent}, {Kleinman}, {Kniazev},
  {Krzesinski}, {Kron}, {Kuropatkin}, {Lebedeva}, {Lee}, {French Leger},
  {L{\'e}pine}, {Levine}, {Lin}, {Long}, {Loomis}, {Lupton}, {Malanushenko},
  {Malanushenko}, {Margon}, {Martinez-Delgado}, {McGehee}, {Monet}, {Morrison},
  {Munn}, {Neilsen}, {Nitta}, {Norris}, {Oravetz}, {Owen}, {Padmanabhan},
  {Pan}, {Peterson}, {Pier}, {Platson}, {Re Fiorentin}, {Richards}, {Rix},
  {Schlegel}, {Schneider}, {Schreiber}, {Schwope}, {Sibley}, {Simmons},
  {Snedden}, {Allyn Smith}, {Stark}, {Stauffer}, {Steinmetz}, {Stoughton},
  {SubbaRao}, {Szalay}, {Szkody}, {Thakar}, {Thirupathi}, {Tucker}, {Uomoto},
  {Vanden Berk}, {Vidrih}, {Wadadekar}, {Watters}, {Wilhelm}, {Wyse}, {Yarger},
  \& {Zucker}}]{segue}
{Yanny}, B., {Rockosi}, C., {Newberg}, H.~J., {et~al.} 2009, \aj, 137, 4377

\bibitem[{{Yong} {et~al.}(2013){Yong}, {Norris}, {Bessell}, {Christlieb},
  {Asplund}, {Beers}, {Barklem}, {Frebel}, \& {Ryan}}]{Yong13}
{Yong}, D., {Norris}, J.~E., {Bessell}, M.~S., {et~al.} 2013, \apj, 762, 27

\bibitem[{{Zhao} {et~al.}(2006){Zhao}, {Chen}, {Shi}, {Liang}, {Hou}, {Chen},
  {Zhang}, \& {Li}}]{lamost}
{Zhao}, G., {Chen}, Y.-Q., {Shi}, J.-R., {et~al.} 2006, \cjaa, 6, 265

\end{thebibliography}

\clearpage

\begin{appendix}
\section{}
\begin{figure}[ht!]
\begin{minipage}{185mm}
\begin{center}
\includegraphics[width=185mm]{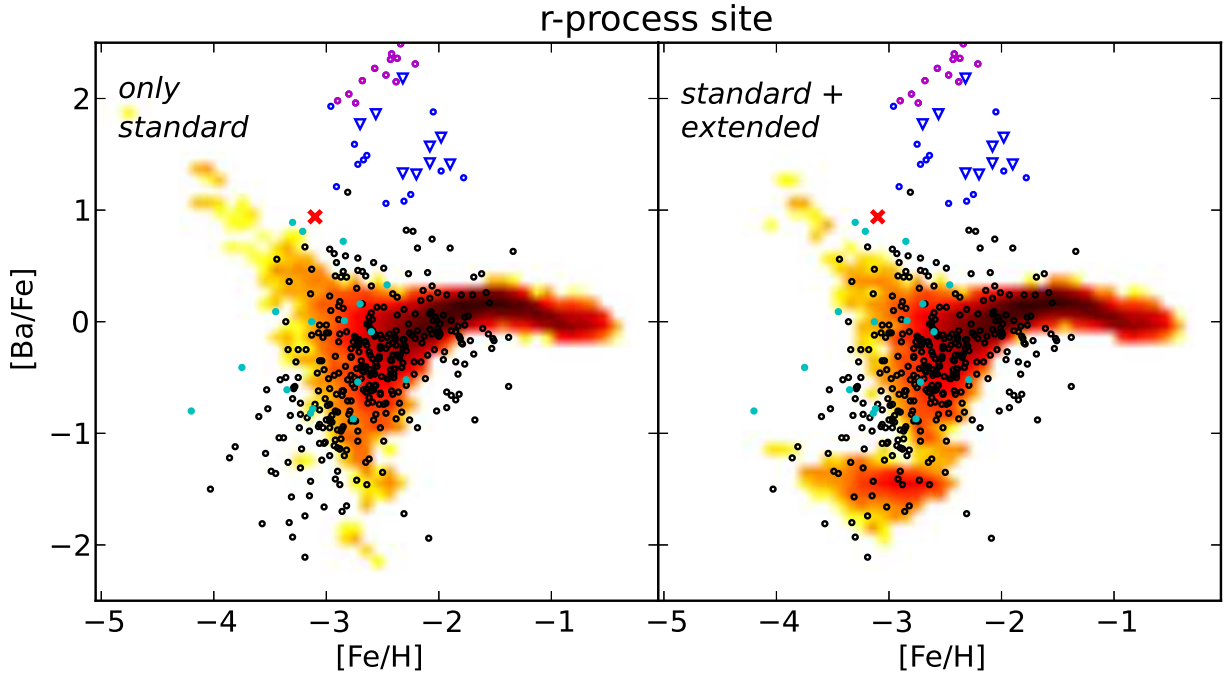}

\caption{ [Ba/Fe] vs [Fe/H] results; on the left the model for an
  r-process contribution only from the ``standard'' site, on the right
  the {\bf r-model}, where both ``standard'' and ``extended''
  contributions are considered.  The density plot is the distribution
  of simulated long-living stars for our models; the density is on a
  log scale, normalized to the peak of the distribution (see bar over
  the Fig. ~\ref{fig1b} for the color scale).  Superimposed, we show
  the abundances ratios for halo stars \citep[data
  from][]{Frebel10}.  The symbols are the same as in
  Fig. ~\ref{fig2}.  }\label{appendix}

\end{center}
\end{minipage}
\end{figure}

We present in Fig. \ref{appendix}, the results of our
  inhomogenous model, with and without the ``extended'' site for the
  r-process.  Thanks to the inhomogeneous model, it is possible to
  highlight that if we use only the ``standard'' r-process site (left
  pannel), most of the stars showing a low [Ba/Fe] cannot be
  reproduced by our model. This problem is not clear from the results
  of the homogeneous model because the simple line representing the
  homogeneous results cannot display that a negligible amount of stars
  are formed by the model in this area.

\end{appendix}

\end{document}